\documentclass[journal,twoside]{IEEEtran}

\usepackage{amsmath}
\usepackage{amssymb}
\usepackage{latexsym}
\usepackage{graphicx}
\usepackage{epsfig}
\usepackage{psfrag}
\usepackage[active]{srcltx} 
\usepackage{color}
\usepackage{setspace}
\usepackage[normal,footnotesize]{caption}
\usepackage{balance}
\usepackage{textcomp}
\usepackage{url}
\usepackage{subfig}

\newtheorem{remark}{\it{Remark}}

\newcommand{\insertonefig}[4]
{
\begin{figure}[htbp]
\begin{center}
\includegraphics[width=#1,clip,keepaspectratio]{#2.pdf}
\end{center}
\caption{#4}
{#3}
\end{figure}
}

\newcommand{\inserttwofigsV}[5]
{
\begin{figure}[h]
\begin{center}$
\begin{array}{c}
\includegraphics[width=#1,clip,keepaspectratio]{#2.pdf} \\
\text{(a)} \\
\linebreak[4] \\
\includegraphics[width=#1,clip,keepaspectratio]{#3.pdf} \\
\text{(b)} \\
\end{array}$
\end{center}
\caption{#5}
{#4}
\end{figure}
}

\newcommand{\addphoto}[1]{\includegraphics[width=1in,height=1.25in,clip,keepaspectratio]{#1.pdf}}

\begin{document}

\title{A Novel Approach to the Statistical Modeling\\of Wireline Channels}

\author{
Stefano~Galli\IEEEmembership{,~Senior~Member,~IEEE}
\thanks{Stefano Galli (sgalli@assia-inc.com) is with ASSIA, Inc. At the time of writing this paper, Dr. Galli was with Panasonic Corporation.}
\thanks{Manuscript submitted Nov.12,~2009. Revised July 28 and Nov. 22,~2010.}%
}

\markboth{To appear in the IEEE Transactions on Communications, 2011}%
{S. Galli \MakeLowercase: A Novel Approach to the Statistical Modeling of Wireline Channels}

\maketitle

\begin{abstract}
We report here that channel power gain and Root-Mean-Square Delay Spread (RMS-DS) in Low/Medium Voltage power line channels are negatively correlated lognormal random variables. Further analysis of other wireline channels allows us to report a strong similarity between some properties observed in power line channels and the ones observed in other wireline channels, e.g. coaxial cables and phone lines. For example, it is here reported that channel power gain and logarithm of the RMS-DS in DSL links are \textit{linearly} correlated random variables. Exploiting these results, we here propose a statistical wireline channel model where tap amplitudes and delays are generated in order to reflect these physical properties. Although wireline channels are considered deterministic as their impulse response can be readily calculated once the link topology is known, a statistical wireline channel model is useful because the variability of link topologies and wiring practices give rise to a stochastic aspect of wireline communications that has not been well characterized in the literature. Finally, we also point out that alternative channel models that normalize impulse responses to a common (often unitary) power gain may be misleading when assessing the performance of equalization schemes since this normalization artificially removes the correlation between channel power gain and RMS-DS and, thus, Inter-Symbol Interference (ISI).
\end{abstract}


\thispagestyle{empty}

\section{Introduction}
\IEEEPARstart{T}{he} issue of channel modeling is of paramount importance as any sensible communications system design must be matched to the particular characteristic of the channel. In particular, the lack of a commonly agreed upon model for the power line (PL) channel has slowed down transceiver optimization and the pursuit of general results in the area of Power Line Communications (PLC). In the last decade, PLC channel modeling efforts have focused on a better understanding of the physics of signal propagation, leading at first to the definition of phenomenological models \cite{ZimDos2002mp} and, more recently, to the definition of deterministic models based on transmission line (TL) theory (see \cite{GalBan2006} and references therein). Although the multipath nature of signal propagation along PLs \cite{ZimDos2002mp} as well as along other TLs \cite{GalWar02} is today well accepted, what is currently lacking in today's models is a realistic statistical characterization of the channel. As a consequence, results available today on the performance of communications systems operating over the PL have limited applicability and cannot be used to draw general conclusions. There are very few contributions attempting to define a statistical model for the PLC channel; furthermore, these contributions are still somewhat incomplete, do not allow physical insight, lack empirical justification and, in some cases, they are difficult to replicate or generalize.

The most widespread PLC channel model is the multipath-based one introduced in \cite{ZimDos2002mp, Barnes1998, philipps:99, ZimDos1999isplc}. According to this model, signal propagation along PL cables is mostly affected by multipath effects that arise from the presence of several branches and impedance mismatches that cause multiple reflections. In this approach, a parametric model of the channel is set forth as a superposition of delayed and attenuated echoes and the model parameters (delay, attenuation, number of paths, etc.) are fitted after measuring the channel. As a consequence, this model cannot be used to compute the transfer function a priori. Although the multipath nature of signal propagation along PLs as well as along other TLs \cite{GalWar02} is today well accepted, there are several disadvantage in this approach. First, it cannot be used to compute the channel transfer function a priori. Second, resonant effects due to parasitic capacitances and inductances as well as particular wiring and grounding practices cannot be explicitly included in the model but only ``phenomenologically'' observed through the initial measurement. Finally, models for the reflection coefficients and for the link topologies are needed to generate the multiple paths.

Using classical two-conductor TL-theory,  Meng et al. complement the multipath model by allowing the a priori computation of the channel model parameters so that preliminary measurements can be avoided \cite{MengChenGuan2002}. This approach requires full knowledge of the link topology, similarly to the frequency domain models based on TL theory. However, since in a time domain model all resolvable paths have to be generated one-by-one, the computational complexity of the multipath models grows with the number of discontinuities and becomes often prohibitively high for the IH case (see, for example, Sect. III.A in \cite{GalBan2006}). For this reason, contributions have recently been focusing on frequency domain deterministic models based on TL-theory \cite{BanGal2001, SartDelo2001, EsmaKsch2003, MengChenGuan2004, BanGalPI2005, GalBanPII2005, GalBan2006, AmirKave2006}. This trend also confirms that a more complete understanding of the physical propagation of communications signals over PLs has been recently emerging.

The major drawback of TL-theory modeling and of the modified multipath one is that the knowledge of the whole topology is required for the computation of the link transfer function. This issue could be by-passed either by creating a set of topologies that can be considered as representative of the majority of topologies that can be found in the field for a specific scenario or by generating randomly a set of statistically relevant transfer functions. In fact, as suggested in \cite{GalBan2006}, the availability of accurate deterministic models based on TL-theory allow us to generate without prior measurements and with good accuracy any topology thus empowering us with the capability of defining hybrid deterministic/Monte Carlo models. However, these solutions are only able to partially alleviate the problem. In fact, the first approach is reminiscent of what has been done in xDSL context with the definition of the ANSI and CSA loops and is of limited applicability to the PL case due to the wide variability of wiring and grounding practices around the world. The second approach still requires a model for the number of branches, their length, their location, etc.

A first attempt to define a model for the generation of random IH topologies has been made by Esmailian et al. \cite{EsmaKsch2003} by using the US National Electric Code to set constraints on the topologies in terms of number of outlets per branch, wire gauges, inter-outlet spacing, etc. On the other hand, a generalization of this approach requires the knowledge of the electric codes of every country. The results presented in \cite{EsmaKsch2003} ignore the effects of ground bonding which is mandatory in the US. However, it is straightforward to include grounding by introducing the ``companion model'' proposed in \cite{BanGalPI2005, GalBanPII2005, GalBan2006} that allows treating under the same formalism both ungrounded and grounded links.

In order to by-pass the dependency of the model on national wiring practices, it was proposed in \cite{Tonello2007} to choose the location of discontinuities according to a Poisson distribution and then generate the impulse response in the time domain by generating echoes one-by-one. Notwithstanding its appealing simplicity, this model lacks physical justification as it is not supported by any empirical observation nor can it be justified by physical considerations. Furthermore, there are no guidelines for setting the parameters of the frequency dependent attenuation or the intensity of the Poisson arrival process in order to simulate a specific scenario. Another attempt towards a statistical modeling of the PLC channel is reported in \cite{TlicZeddMoulTPDI2008}. Nine classes of channels and their respective transfer functions are defined and it is observed that peak and notch widths, heights, and numbers of a smoothed version of the measured transfer functions are fitted by Rayleigh, triangular, and Gaussian distributions, respectively. However, there is very little physical insight about why such distributions arise and one may wonder if these distributions are simply an artifact of the chosen classification procedure. Furthermore, the operation of smoothing of the transfer function applied in \cite{TlicZeddMoulTPDI2008} is questionable since it changes considerably the degree of frequency selectivity of the channel and, thus, the distortion that is introduced.

In this paper, we address the above mentioned lack of a physically meaningful statistical characterization of the PLC channel by reporting statistical results that shed light on important properties of the Low Voltage (LV) and Medium Voltage (MV) PLC channels. In particular, we confirm here on the basis of channel measurements that channel power gain and RMS-DS of LV/MV PLC channels are correlated lognormal random variables - initial results were reported for the first time in \cite{Galli2009lognormal}. This paper also reports for the first time that the above mentioned correlation is also a characteristic of other wireline communications channels. For example, it has been found that channel power gain and the logarithm of the RMS-DS in DSL links are \textit{linearly} correlated random variables.

An important consequence of this property is that performance results obtained on the basis of channel models that contain some sort of channel gain normalization should be accepted with the grain of salt as channel gain normalization removes the correlation between channel gain and RMS-DS and, thus, ISI. For example, these models would not be able to capture the physical property that channels characterized by a high attenuation are also characterized by severe ISI. Similar considerations may also be made in the wireless context as several researchers have reported empirical results confirming that the correlation between channel gain and RMS-DS is sometimes observable in radio channels (see Sect. \ref{sec.UniversalWireless} for more details).

On the basis of these results, we here propose a new approach to wireline statistical channel modeling where the correlation between gains and RMS-DS is imposed by design. We also show how the proposed approach allows us to define a very simple two-tap statistical channel model that, despite its simplicity, is capable of producing realistic capacity Cumulative Distribution Function (CDF) plots. The proposed statistical model allows us to properly replicate the variability of link topologies and wiring practices that actually confer a truly stochastic aspect to deterministic TL-based channels. The availability of statistical channel models will aid in gaining a better understanding of the range and coverage that PLC solutions can achieve, a necessary prerequisite when deploying PLC equipment in the field - especially for Smart grid applications \cite{Link:GalliScaglioneWang2011ProcIEEE}.

The paper is organized as follows. In Sect. \ref{sec.OverviewStats}, we report statistical properties of LV/MV PLC channels. In Sect. \ref{sec.Universality}, we point out the commonality of channel gain/RMS-DS correlation in several wireline channels and also confirm the general validity of the lognormal model for channel gains. The implications of the found correlation property from an equalization perspective are discussed in Sect. \ref{sec.ISImitigation} for both single an multi-carrier schemes. The proposed statistical channel modeling approach is outlined in Sect. \ref{sec.GeneralModel}, where we also discuss the advantages of a simple two-tap channel model. Simulation results are presented in Sect. \ref{sec.Simulations} and finally conclusive remarks are drawn in Sect. \ref{sec.Conclusions}.

\section{Statistical Results for LV and MV\\Power Line Links}\label{sec.OverviewStats}
In this section, we report statistical results obtained on the basis of measured PLC channels. For the LV case, we report channel measurements of US sub-urban and urban homes. The sub-urban measurements reported here pertain to a set of channels measured by the HomePlug Powerline Alliance that have been made available online \cite{ITU:PLCdata:NIPP}. The set contains 60 forward and reverse PLC transfer functions taken from six different homes of various sizes and age in the US - for more details see \cite{Mahony06}. For the US urban case, we use 40 forward and reverse PLC transfer functions collected by Panasonic in five US apartment buildings (Multiple Dwelling Units). In all cases, we have considered the 1.8-30 MHz band and only the forward links (the reverse links yielded approximately the same values of average gain and RMS-DS as predicted by theory \cite{BanGal2001sym}). The MV PL links have been measured in a 3-phase MV underground network. A single drive point was chosen, B-phase at a transformer, and then many measurement points were chosen at various locations, progressing in distance from 21.3 m (70 ft) out to about 365.8 m (1200 ft) away from this central feed point. This data includes going through features such as junction boxes, taps transformers, and phase-to-phase cross-coupling. In all reported measurements, the considered channel bandwidth is [2-40] MHz. Two different types of coupling methods were used for the MV measurements, both inductive coupling on the center conductor and inductive coupling on the concentric neutral.

In the following Subsections, we will use the following definitions:
\begin{itemize}
\item Impulse response: $h_i \triangleq h(t=iT_S)$, $i=0, 1, 2, \ldots, L-1$, obtained sampling at rate $F_S=1/T_S$ the continuous time impulse response $h(t)$. The channel memory is $L$-1, and there are $L$ non zero taps. Channel tap amplitudes and gains at delay $\theta_i=iT_S$ are $h_i$ and $P(\theta_i)=\lvert h_i \rvert^2$, $i=0, 1, \ldots, L-1$, respectively.
\item Transfer Function: $H_i \triangleq \lvert H_i \rvert e^{j\phi_i}$, $i=0, 1, 2, \ldots, N-1$, obtained as the $N$-point DFT of the discrete impulse.
\end{itemize}

On all data sets we have tested for lognormality at the 5\% significance level using the following tests: Jarque-Bera, Shapiro-Wilk/Francia\footnote{$^)$~The Shapiro-Wilk test is performed when data samples are platykurtic, while the Shapiro-Francia test is performed when they are leptokurtic.}, Lillilifors, Anderson-Darling, and Chi-square. Furthermore, we also used the Kolmogorov-Smirnov test to verify that two data sets come from the same (normalized) distribution.

\subsection{Average Power Channel Gain}\label{sec.AveChanGains}
The channel power gain $\overline{G}$ can be expressed as:
\begin{equation}\label{eq.GainDef}
    \overline{G} = \sum_{n=0}^{L-1} |h_n|^2  = \frac{1}{N} \sum_{i=0}^{N-1} |H_i|^2 =  \sum_{n=0}^{L-1} P(\theta_i),
\end{equation}
where the term ``average'' is used here to differentiate this power gain from the individual channel power gain $|H_i|^2$ at a certain frequency or sub-carrier. It is possible to justify the lognormality (normality) of the average channel gains $\overline{G}$ ($\overline{G}_{dB}=10log(\overline{G})$) of PLC channels using several arguments related to either the multipath nature of signal propagation or to TL modeling based on cascaded two-port networks. Actually, these arguments hold not only for the PLC case but also for the general case of TL-based channels.

Considering signal propagation along TLs as multipath-based \cite{ZimDos2002mp, GalWar02}, channel distortion is present at the receiver due not only to the low pass behavior of the cable but also to the arrival of multiple echoes caused by successive reflections of the propagating signal generated by mismatched terminations and impedance discontinuities along the line. This is a general behavior and is independent of the link topology or, in the case of PLs, of the presence of grounding \cite{GalBan2006}. According to this model, the transfer function is \cite{ZimDos2002mp}:
\begin{equation}\label{eq.Hfmp}
    H(f) = \sum_{i=0}^{N_{paths}-1} g_i(f) e^{-\alpha (f) v_p \theta_i} e^{-j2 \pi f \theta_i}
\end{equation}
where $g_i(f)$ is a complex number generally frequency dependent that depends on the topology of the link, $\alpha (f)$ is the attenuation coefficient which takes into account both skin effect and dielectric loss, $\theta_i$  is the delay associated with the $i$-th path, $v_p$ is velocity of propagation along the PLC cable, and $N_{paths}$ is the number of non-negligible paths. Similarly, we can write in the time domain:
\begin{equation}\label{eq.htmp}
    h(t) = \sum_{i=0}^{N_{paths}-1} e_{ep}^{(i)}(t-\theta_i)
\end{equation}
where $e_{ep}^{(i)}(t)=FT^{-1} \left[ g_i(f) e^{-\alpha (f) v_p \theta_i} \right]$ is the signal propagating along the $i$-th path. The amplitude and shape of the $i$-th path signal are a function of the reflection coefficients $\rho^{(i)}$ and the transmission coefficients $\xi^{(i)}=(1+\rho^{(i)})$ associated to all the impedance discontinuities encountered along the $i$-th path, and of the low-pass behavior of the channel in the absence of multipath (for analytical expressions of $\rho^{(i)}$ and $\xi^{(i)}$, see \cite{GalBan2006} for the case of forward traveling signal paths and \cite{GalWar02} for the case of backward traveling echo paths). Thus, the path amplitudes are a function of a cascade (product) of several random propagation effects and this is a condition that leads to lognormality in the central limit since the logarithm of a product of random terms becomes the summation of many random terms. Since lognormality is preserved under power, path gains are lognormally distributed as well. Finally, since the sum of independent or correlated lognormal random variables is well approximated by another lognormal distribution \cite{MehMolWu06}, we can finally state that also average channel gains $\overline{G}$ are lognormally distributed.

When considering TL-based channel models based on cascaded two-port networks and transmission (or ABCD) matrices, each section of cable, each bridged tap, the mains breaker box and the eventually present grounding are modeled as a two-port network and are represented by a two-by-two transmission matrix \cite{GalBan2006}. The transfer function of the overall TL link is equal to the reciprocal of a linear combination of the four elements (A, B, C, D) of the overall transmission matrix which is calculated exploiting the chain rule, i.e. the overall transmission matrix is equal to the product of the transmission matrices of all cable sections, bridged taps, etc. Thus, the overall transmission matrix is obtained as the product of many matrices which can be considered random when considering a generic topology. There is an extensive literature relating lognormal distributions to the product of multiple random matrices \cite{OlivPetri96}, so again lognormality of individual and average channel gains arises again. The above theoretical considerations hold for any PLC network regardless of the frequency band of operation.

All statistical tests confirmed lognormality of $\overline{G}$ at the 5\% significance level for all IH data sets presented here. Table \ref{tab.PLCstats} reports notable statistical values for the average channel gains expressed in dB. We remark the large excursion of channel attenuation encountered in the PLC channel, around 50 dB. This excursion is much wider than that generally encountered in IH phonelines (PH) or in IH coaxial cables (CX) which is around 20 dB.

For the MV case, lognormality tests at the 5\% significance level have failed when all 59 data points were used. By removing six suspected outliers using boxplot analysis, channel gains in dB have passed all considered normality tests with a minimum \textit{pValue} of 0.4. The identified outliers corresponded to cases where channel attenuation was very high ($>65~dB$) and link capacity was zero.

\begin{table*}[htbp]
\centering
\caption{Statistical values in dB and in $\mu s$ of the measured channel attenuation $\overline{A}_{dB} = - \overline{G}_{dB}$ and the measured RMS-DS $\sigma_\tau$ of US IH LV and MV underground links. In the case of sub-urban homes, we also report statistical values of the conditional distribution of RMS-DS when $\overline{A}_{dB}>45$ dB. We have used 60, 40 and 59 measured channels for the IH sub-urban, IH urban, and MV cases, respectively.}
\label{tab.PLCstats}
{
\renewcommand{\arraystretch}{1.2}

\begin{tabular}{l|ccc|cc|cc|}
\cline{2-8}\cline{2-8}
 ~ & \multicolumn{3}{|c|}{\textbf{IH PLC - sub-urban}} & \multicolumn{2}{|c|}{\textbf{IH PLC - urban}} & \multicolumn{2}{|c|}{\textbf{MV PLC}} \\ [0.5ex]
 ~ & $\overline{A}_{dB}$ & $\sigma_\tau$ $(\mu s)$ & $\sigma_\tau|_{\overline{A}_{dB}>45}~(\mu s)$ & $\overline{A}_{dB}$ & $\sigma_\tau$ $(\mu s)$  &  $\overline{A}_{dB}$ & $\sigma_\tau$ $(\mu s)$\\ [0.5ex]
\hline\hline
Min                             & 19.7   & 0.10    &  0.2    & 14.5  & 0.11 &  10.2  &   0.17 \\
Max                             & 68.1   & 1.73    &  1.7    & 65.1  & 0.47 &  82.5  &   0.78 \\
Mean ($\mu$)                    & 48.9   & 0.52    &  0.6    & 41.5  & 0.23 &  45.2  &   0.52 \\
Std. Dev. ($\sigma$)            &  9.8   & 0.28    &  0.3    & 13.4  & 0.09 &  13.2  &   0.15 \\
Kurtosis                        &  3.9   & 7.60    &  7.7    & 2.3   & 3.82 &  4.9   &   2.79 \\
Skewness                        & -0.6   & 1.70    &  1.8    & -0.4  & 0.87 &  0.7   &  -0.60 \\
50\%-Percentile                 & 49.4   & 0.46    &  0.5    & 44.3  & 0.23 &  43.7  &   0.55 \\
90\%-Percentile                 & 60.4   & 0.94    &  1.0    & 58.1  & 0.34 &  61.1  &   0.68 \\ [1ex]
\hline
\end{tabular}
}
\end{table*}

\subsection{RMS Delay Spread}\label{sec.RMSDSdef}
The RMS-DS $\sigma_\tau$ is given by the following relationship:
\begin{equation}\label{eq.RMSDSdef1}
\sigma_\tau \triangleq \sqrt{\mu_\tau^{(2)}-(\mu_\tau)^2} = T_S \sqrt{\mu_0^{(2)}-(\mu_0)^2}
\end{equation}
\begin{eqnarray}\label{eq.RMSDSdef2}
\mu_\tau  &=& \frac{\sum_{i=0}^{L-1} \theta_i P(\theta_i)}{\sum_{i=0}^{L-1} P(\theta_i)}= T_S \mu_0 \\
\mu_\tau^{(2)} &=& \frac{\sum_{i=0}^{L-1} \theta_i^2 P(\theta_i)}{\sum_{i=0}^{L-1} P(\theta_i)}= (T_S)^2 \mu_0^{(2)}
\end{eqnarray}
where $\mu_0$ and $\mu_0^{(2)}$ are defined as:
\begin{equation}\label{eq.RMSDSnormdef}
\mu_0   = \frac{\sum_{i=0}^{L-1} i \lvert h_i \rvert^2}{\sum_{i=0}^{L-1} \lvert h_i \rvert^2}, \hspace{1cm} \mu_0^{(2)} = \frac{\sum_{i=0}^{L-1} i^2 \lvert h_i \rvert^2}{\sum_{i=0}^{L-1} \lvert h_i \rvert^2} .
\end{equation}

Table \ref{tab.PLCstats} reports notable statistical values for the RMS-DS expressed in $\mu s$. The RMS-DS of sub-urban US homes exhibits very high kurtosis (leptokurtic behavior), thus indicating that its distribution is more outlier-prone than the Gaussian one which has a kurtosis equal to $3$. We have also considered the two sets of RMS-DS conditional to the channel attenuation being larger or smaller than $45$ dB. The RMS-DS statistical values for the case when the attenuation is larger than $45$ dB are also shown in Table \ref{tab.PLCstats} and are useful to generate channel realizations in leptokutic scenarios as the IH PLC sub-urban one - see Step 3 in the channel generation procedure given in Sect. \ref{sec.GeneralModel}.

Among the 60 measured responses, 59 channels exhibited an RMS-DS below $1.3 \mu s$. One response only exhibited an RMS-DS value of $1.7 \mu s$. On the other hand, data collected in US urban locations show kurtosis closer to the Gaussian case. Similar results can be found in \cite{LiuFlint1999, EsmaKsch2003} where values of RMS-DS under $1 \mu s$ have been reported in similar settings. Similarly, the RMS-DS of the measured MV PL links were all under $1~\mu s$. We can then state that the RMS-DS of a PL link is much smaller than traditionally reported and, for the considered scenarios, it is below $1 \mu s$ in most cases. The symbol interval of commercially available PLC modems is $8.192 \mu s$ (case of Wavelet-OFDM \cite{GalKogKod2008}, \cite{GalLog2008}) or $40.96 \mu s$ in the case of windowed OFDM with Cyclic Prefix (CP) \cite{AfkLatYon05}. Therefore, in the vast majority of cases, the RMS-DS of the PLC channel is less than 20\% of the baud interval. We point out that several published papers report that typical IH ``delay spreads'' are in the order of few microseconds, e.g. see \cite{AbadBadeBlas2003, GotRapDos2004, AfkLatYon05}. For example, it is reported in \cite{AfkLatYon05} that measured IH ``delay spreads'' are in the order of $2-3 \mu s$, with some exceptional cases of delay $5 \mu s$. However, most papers do not generally calculate the actual RMS-DS and what is actually reported is the duration of the impulse response that includes a percentage $\alpha$ of the total impulse response energy - for example, a value of $\alpha=0.95$ has been used by Afkhamie et al. in \cite{AfkLatYon05}.

From the lognormality of path gains $P(\theta_i)$, it should not be surprising that also the RMS-DS is lognormally distributed since lognormality is preserved under scaling and power, and quasi-preserved under addition. This was verified for the case of IH PL links, where all considered lognormality tests succeeded at the 5\% significance levels - including the two conditional cases introduced for the IH PL sub-urban case. Although the Kolmogorov-Smirnov test confirmed that $\overline{G}$ and $log(\sigma_\tau)$ belong to the same distribution, lognormality was not confirmed by all tests for the RMS-DS of the MV PL data set. However, after removing ten potential outliers corresponding to the smallest RMS-DS values, lognormality was confirmed by all statistical tests with a \textit{pValue} ranging between 0.2 and 0.5. The fact that small values of RMS-DS cause a deviation from lognormality should not be surprising as confirmed by Figure 6 in \cite{Galli2009lognormal} where the quantile-quantile plot of $log(\sigma_\tau)$ versus normal quantiles shows a small deviation from the linear trend at low RMS-DS. Further confirmation of this behavior on the basis of simulation results has also been recently reported in \cite{ToneVers2010BottomUp2} for the RMS-DS of short indoor loops.

\insertonefig{8.8cm}{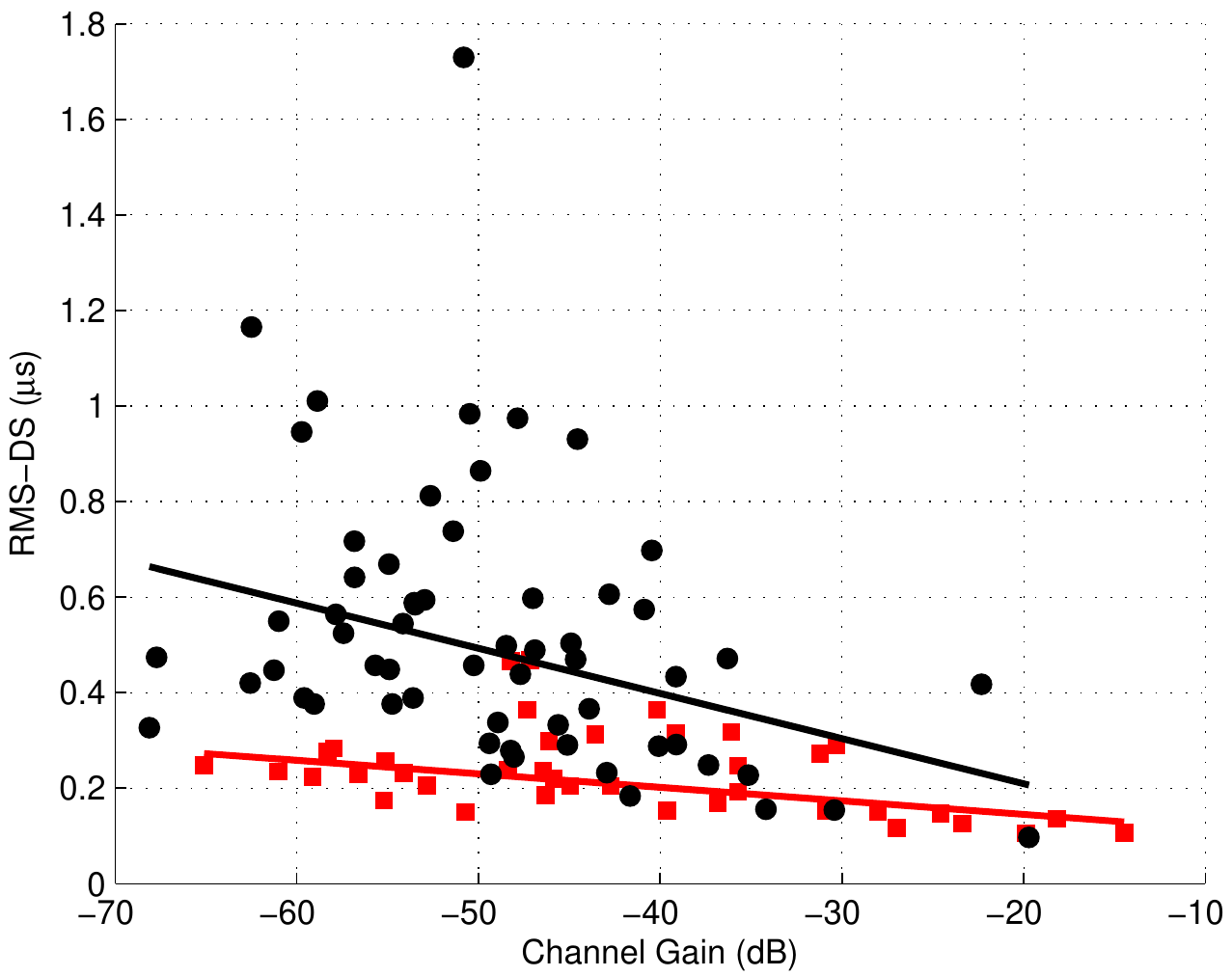}{\label{fig.PLCScatter}}
{Scatter plot of measured indoor power line channels with least squares trend lines: US sub-urban black circle) and US urban (red square) homes.}

\insertonefig{8.8cm}{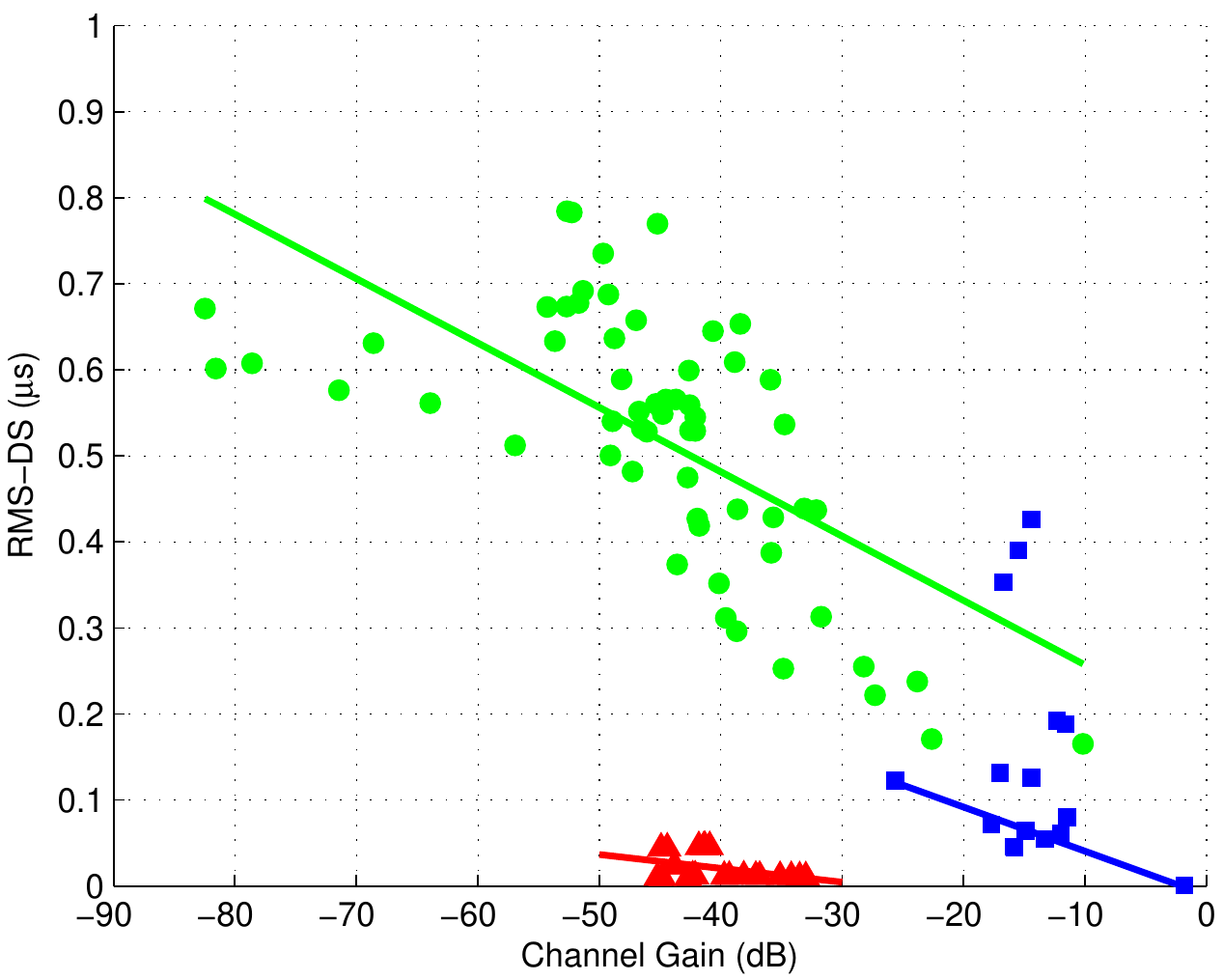}{\label{fig.AllScatter}}
{Scatter plot of measured and simulated channels with least squares trend lines: measured underground MV PLC links (green circles), simulated indoor phone lines (blue squares), and simulated indoor coax links (red triangles).}

\section{On the Relationship Between Channel Gains and RMS-DS:A Universal Property of Wireline Channels} \label{sec.Universality}
The early results reported in \cite{Galli2009lognormal, Galli2010TwoTaps} are confirmed here for the case of IH links (see Figure \ref{fig.PLCScatter}) as well as for other wireline channels like the UG MV, IH CX, IH PH, and DSL ones (see Figures \ref{fig.AllScatter} and \ref{fig.DSLplots}). The scatter plots are shown together with a trend line given by the equation below:
\begin{equation}\label{eq.PLCrobregr}
    \sigma_{\tau,\mu s} = \alpha \cdot \overline{G}_{dB} +\beta
\end{equation}
where $\sigma_{\tau,\mu s}$ is the RMS-DS in $\mu s$ and $\overline{G}_{dB}$ is the average channel gain in dB. In some cases, a higher correlation was found between channel gains and the logarithm of the RMS-DS so that in some cases we also report the line equation that ties $log(\sigma_{\tau,\mu s})$ and $\overline{G}_{dB}$:
\begin{equation}\label{eq.PLCrobregrLOG}
    log(\sigma_{\tau,\mu s}) = \theta \cdot \overline{G}_{dB} + \zeta
\end{equation}

Parameters $\alpha$, $\beta$, $\theta$ and $\zeta$ are calculated using a robust iteratively reweighted least squares algorithms with a bi-square weighting function that assigns a smaller weight to data points farther from model prediction. All robust regression lines exhibit a negative slope clearly confirming that channel gains and RMS-DS (and its logarithm) are negatively correlated. Therefore, channels with large RMS-DS (severe ISI) are also characterized by small channel gains (large attenuation and, thus, also low SNR for fixed transmit power), and vice versa.

Correlation between channel gain and RMS-DS appear to be a property common to all examined wireline channels. However, lognormality was not always confirmed by all statistical tests. This probably depends on the fact that in some cases too few data samples were available to give an accurate answer.

Tables \ref{tab.PLCstats} and \ref{tab.ALLstats} report notable statistical values for the average channel gains (dB) and the RMS-DS ($\mu s$) of IH PL (urban and sub-urban), UG MV PL, IH CX, IH PH, and DSL (ANSI and CSA). In the next Sub-Sections, we will investigate in detail these cases and report the values of the parameters in eqs. \eqref{eq.PLCrobregr} and \eqref{eq.PLCrobregrLOG} for each scenario. Furthermore, the same statistical tests described in Sect. \ref{sec.OverviewStats} have been carried out on all data sets to investigate the distribution of gains and RMS-DS. The results presented in this Section are summarized in Table \ref{Tab.Summary}, together with the results presented in the previous Section.

\begin{table*}[htbp]
\centering
\caption{Statistical values in dB and in $\mu s$ of channel attenuation $\overline{A}_{dB} = - \overline{G}_{dB}$ and RMS-DS $\sigma_\tau$ of simulated CX, PH, and DSL links. We have used a data set of 22, 16, 13, and 10 channels for the IH CX, PH, DSL-ANSI, and DSL-CSA cases, respectively.}
\label{tab.ALLstats}
{
\renewcommand{\arraystretch}{1.2}
\begin{tabular}{l|cc|cc|cc|cc|}
\cline{2-9}\cline{2-9}
 ~ & \multicolumn{2}{|c|}{\textbf{IH CX}} & \multicolumn{2}{|c|}{\textbf{IH PH}} & \multicolumn{2}{|c|}{\textbf{DSL-ANSI}} & \multicolumn{2}{|c|}{\textbf{DSL-CSA}}  \\ [0.5ex]
 ~ & $\overline{A}_{dB}$ & $\sigma_\tau$ $(ns)$  & $\overline{A}_{dB}$ & $\sigma_\tau$ $(\mu s)$ & $\overline{A}_{dB}$ & $\sigma_\tau$ $(\mu s)$  & $\overline{A}_{dB}$ & $\sigma_\tau$ $(\mu s)$ \\ [0.5ex]
\hline\hline
Min                             &  33.0  & 10   &  1.8   & 0.001  & 58.6  & 13.2  & 50.8  & 5.2  \\
Max                             &  45.2  & 46.1 & 25.6   & 0.43   & 65.2  & 27.2  & 57  & 10.9  \\
Mean ($\mu$)                    &  40.3  & 21.6 & 14.4   & 0.15   & 60.1  & 18.0  & 53.1 & 7.1  \\
Std. Dev. ($\sigma$)            &   3.9  & 15.3 & 4.8    & 0.13   & 2.0  & 4.3    & 1.8    & 1.6  \\
Kurtosis                        &   2.1  & 1.92 & 7.3    & 3.4    & 3.6  & 3.0    & 4.0    & 5.9  \\
Skewness                        &  -0.6  & 0.94 & -0.4   & 1.2    & 1.0  & 0.86   & 1.1   & 1.6  \\
50\%-Perc.                      &  41.4  & 11.7 & 14.7  & 0.1    & 60.4  & 17.3  & 52.7  & 6.6  \\
90\%-Perc.                      &  44.8  & 45.9 & 17.6  & 0.4    & 63.7  & 23.8   & 56.2  & 9.8  \\ [1ex]
\hline
\end{tabular}
}
\end{table*}

\subsection{In-Home Power Line Links}
The correlation coefficients between average channel gains and RMS-DS are -0.4 and -0.5 for the US sub-urban and urban cases, respectively. The robust regression parameters $\alpha$ and $\beta$ have the following values:
\begin{itemize}
  \item $\alpha=-0.094~~\mu s/dB$; $\beta=0.02  \mu s$~~~(US sub-urban)
  \item $\alpha=-0.0028~\mu s/dB$; $\beta=0.089 \mu s$~~~(US urban)
\end{itemize}
The higher kurtosis of the RMS-DS of US sub-urban homes can be immediately verified by noting the dispersion of data points around the trend line.

We have observed that the correlation coefficients slightly increase when the logarithm of the RMS-DS is considered. Specifically, the correlation coefficients between $\overline{G}_{dB}$ and $log(\sigma_{\tau,\mu s})$ are -0.5 and -0.6 for the US sub-urban and urban cases, respectively. The robust regression parameters $\theta$ and $\zeta$ have the following values:
\begin{itemize}
  \item $\theta=-0.027~~\mu s/dB$; $\zeta=-2.12 \mu s$~~~(US sub-urban)
  \item $\theta=-0.0167~\mu s/dB$; $\zeta=-2.26 \mu s$~~~(US urban)
\end{itemize}

As already mentioned in Sect. \ref{sec.OverviewStats}, lognormality of $\overline{G}_{dB}$ and $\sigma_\tau$ has been confirmed for both the sub-urban and urban cases by all statistical tests.

\subsection{Underground MV Power Line Links}

The correlation coefficient between average channel gain and RMS-DS is -0.65 and the robust regression parameters $\alpha$ and $\beta$ have the following values:
\begin{itemize}
  \item $\alpha=-0.0075~\mu s/dB$;~~$\beta=0.183 \mu s$~~~(MV PL)
\end{itemize}

For this case, we have not found that the correlation coefficient increases when the the logarithm of the RMS-DS is considered. As mentioned in Sect. \ref{sec.OverviewStats}, lognormality of $\overline{G}_{dB}$ and $\sigma_\tau$ has been confirmed provided that suspect outliers were eliminated.

\subsection{In-Home Coaxial Links}
For the IH CX  case, we have used a very accurate software IH CX channel simulator. The impulse responses of 22 typical IH CX topologies as the ones presented in \cite{Book:Chen2003, STD:HomePNA-2007} were generated and the observed correlation between $\overline{G}_{dB}$ and $\sigma_\tau$ was found to be -0.4.

Statistical values for gains {dB} and RMS-DS ($\mu s$) are reported in Table \ref{tab.ALLstats} and the robust regression parameters are:
\begin{itemize}
  \item $\alpha=-0.0016~\mu s/dB$;~~$\beta=-0.044 \mu s$
\end{itemize}
For this case, we have not found that the correlation coefficient increases when the the logarithm of the RMS-DS is considered.

All employed normality tests confirmed at the 5\% significance level the lognormality of channel gains, with a minimum \textit{pValue} of 0.1. However, only the Jarque-Bera and Chi-square tests confirmed lognormality of RMS-DS even though the Kolmogorov-Smirnov test confirmed that $\overline{G}_{dB}$ and $log(\sigma_\tau)$ come from the same distribution. Empirical data is necessary to assess conclusively the distribution of RMS-DS in IH CX links.

\subsection{In-Home Phone Links}
For the IH PH case, we have used a very accurate software channel simulator. We have computed the impulse responses of 16 IH PH wiring topologies. Specifically, we have used the ten topologies given in the ITU-T G.9954 Recommendation \cite{STD:HomePNA-2007} plus additional 6 obtained by terminating on an open termination all the bridged taps in the G.9954 topologies - the reason for doing so was to increase the number of available data samples. The correlation between $\overline{G}_{dB}$ and $\sigma_\tau$ was found to be -0.2, which is much lower than what has been observed in the other wireline channels analyzed here. However, when the logarithm of the RMS-DS is considered then the correlation coefficient grows considerably to -0.6. The larger correlation between channel gains and the logarithm of the RMS-DS found in IH PH has also been encountered in outdoor phone lines - see Sect. \ref{Sec.UniversalityDSL} on DSL and Table \ref{Tab.Summary}.

The robust regression parameters are:
\begin{itemize}
  \item $\alpha=-0.005~\mu s/dB$;~~$\beta=0.054 \mu s$~~~(Indoor PH)
\end{itemize}
\begin{itemize}
  \item $\theta=-0.007~\mu s/dB$;~~$\zeta=-2.27 \mu s$~~~(Indoor PH)
\end{itemize}

Lognormality of channel gains has not been confirmed as the considered statistical tests gave mixed results, some tests confirmed lognormality and some did not. As far as the RMS-DS, lognormality was confirmed after removing one single outlier (the smallest value equal to 1 ns). Although the Kolmogorov-Smirnov test confirmed that $\overline{G}_{dB}$ and $log(\sigma_\tau)$ come from the same distribution, additional data is necessary to assess conclusively the distribution of the gains of IH PH links.

\begin{remark}
Although it may be possible to find in the field the phone topologies reported in Recommendation ITU-T G.9954, it is also important to point out that those topologies were chosen with the explicit goal of ``stressing'' equalization performance tests. As a consequence, the RMS-DS values of the channels based on these topologies tend to be larger than what is usually found in the field and the operation of opening all the bridged tap terminations has also exacerbated this trend. Therefore, the clearly visible outliers present in the scatter plot of Figure \ref{fig.AllScatter} should be considered really rare to find in the field.
\end{remark}

\subsection{DSL Links} \label{Sec.UniversalityDSL}
Figure \ref{fig.DSLplots} reports the $\overline{G}_{dB}-\sigma_{\tau}$ scatter plot obtained for two types of DSL loops (CSA and ANSI) and for a straight AWG26 cable with no bridged taps and length ranging between $304.8~m$ ($1~kft$) and $5.5~km$ ($18~kft$). It is interesting to note that large values of RMS-DS are generated even when there is no multipath (straight AWG26 lines). In this case, all ISI is generated by the low-pass behavior of cables which is more pronounced in longer cables.

The plot in Figure \ref{fig.DSLplots} also shows that the ten CSA and the thirteen ANSI topologies give rise to the channel gains and RMS-DS similar to those of straight AWG26 cables of length ranging between 2.3-3.4 km (7.5-11 kft) and 3.7-5.3 km (12-17.5 kft), respectively. The correlation between $\overline{G}_{dB}$ and $\sigma_\tau$ was found to be the highest of all investigated cases: -0.97 for ANSI loops and -0.95 for CSA loops.

Statistical values for gains {dB} and RMS-DS ($\mu s$) are reported in Table \ref{tab.ALLstats}, columns five and six. The robust regression parameters are given below:
\begin{itemize}
  \item $\alpha=-0.833~~\mu s/dB$; $\beta=-37 \mu s$~~(DSL-CSA)
  \item $\alpha=-2.1~\mu s/dB$; $\beta=-109  \mu s$~~~~(DSL-ANSI)
\end{itemize}

An interesting thing to note in Figure \ref{fig.DSLplots}.(a) is the similarity between the exponential-like relationship between gains and RMS-DS in straight AWG26 cables and the exponential outer-bound for the RMS-DS versus path-loss (including shadowing) reported by several authors for the cellular environment \cite{FeueBlacRappa1994, SousJovaDaig1994, GreErcYeh97}. Furthermore, the exponential relationship between $\overline{G}_{dB}$ and $\sigma_{\tau}$ becomes a \textit{linear} relationship (correlation coefficient of 0.99) between $\overline{G}_{dB}$ and $log(\sigma_{\tau})$ as shown in Figure \ref{fig.DSLplots}.(b). Linear correlation between $\overline{G}_{dB}$ and $log(\sigma_{\tau})$ extends to both short and long DSL links, whereas both CSA and ANSI loops show a good correlation also between $\overline{G}_{dB}$ and $\sigma_{\tau}$. The parameters expressing this linear relationship are given below:
\begin{itemize}
  \item $\theta=-5.6~~\mu s/dB$; $\zeta=-0.139 \mu s$~~~(DSL-AWG26)
  \item $\theta=-3.85~~\mu s/dB$; $\zeta=-0.109 \mu s$~~(DSL-CSA)
  \item $\theta=-3.81~~\mu s/dB$; $\zeta=-0.11 \mu s$~~~(DSL-ANSI)

\end{itemize}

Lognormality of channel gains and RMS-DS has been confirmed by all tests for the ANSI, and CSA. For the straight AWG26 case, lognormality of channel gains and RMS-DS has been confirmed by all statistical tests with the exception of the Shapiro-Wilk that returned a $pValue$ of 0.045 for the RMS-DS case. The same tests confirmed also normality of all the data sets, thus additional data is needed to assess conclusively the distribution of channel gains and RMS-DS.

\inserttwofigsV{8.8cm}{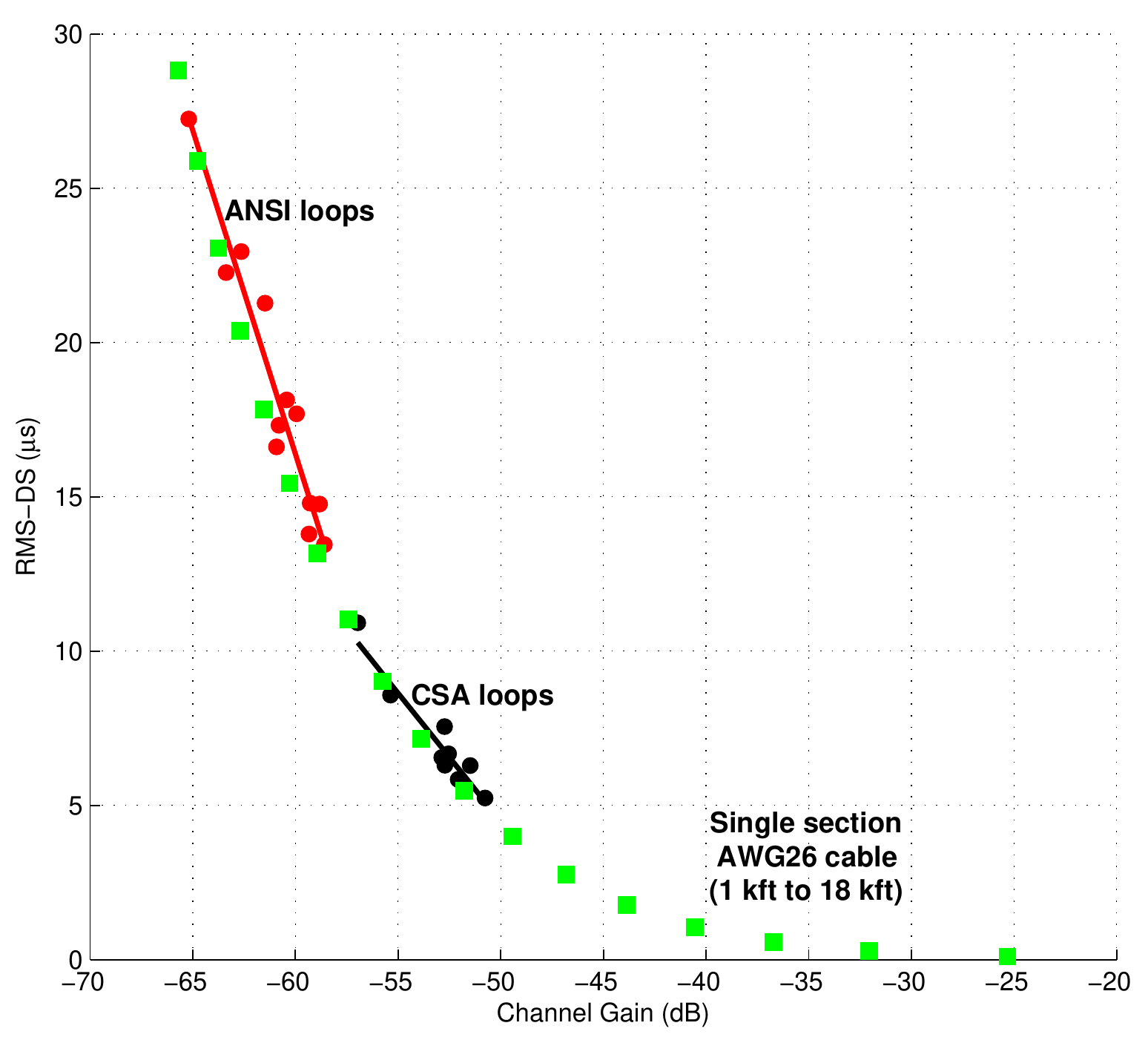}{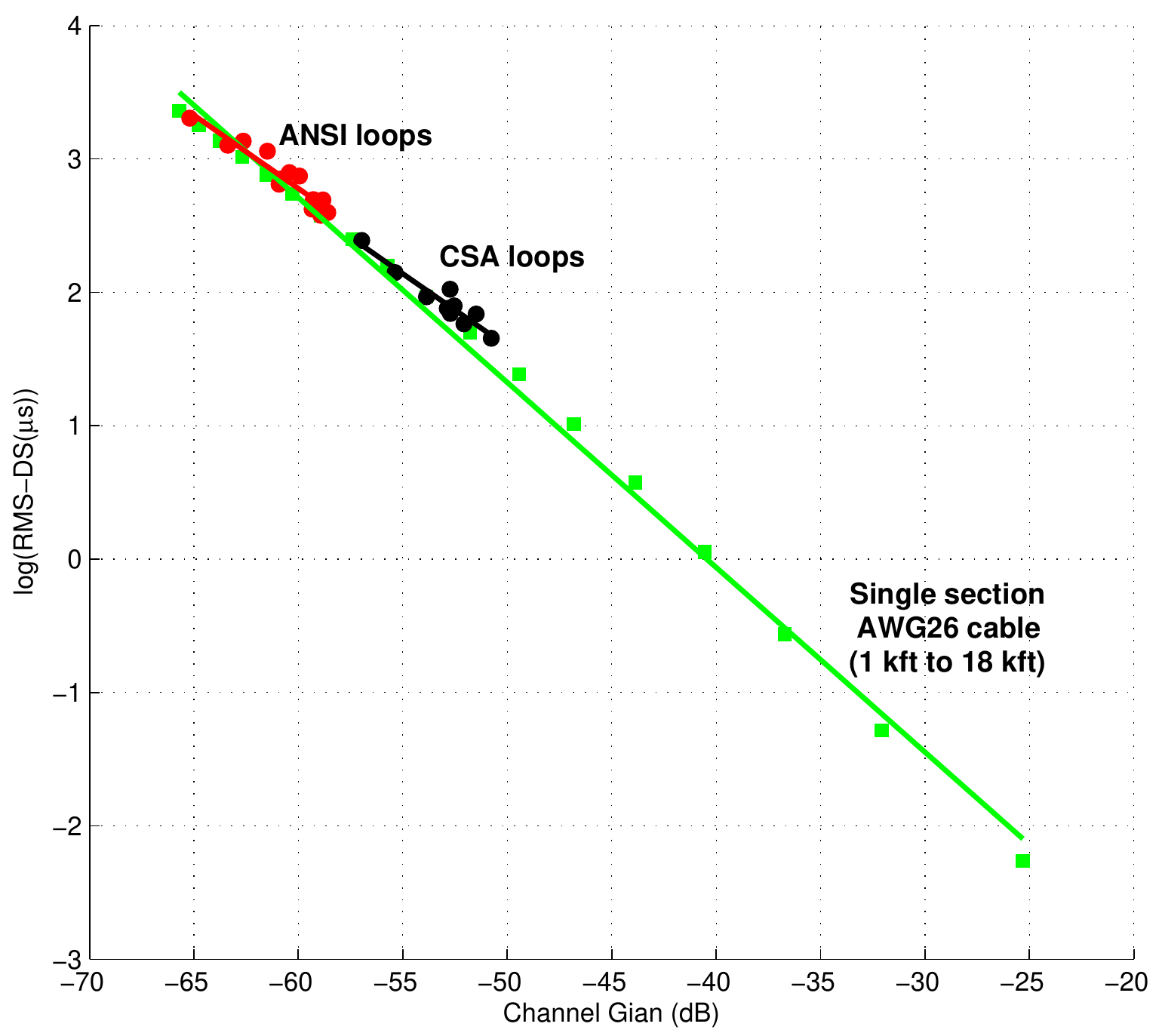}{\label{fig.DSLplots}}
{Scatter plot of simulated channels with least squares trend lines in the: (a) RMS-DS vs channel gains; (b) log(RMS-DS) vs channel gains. (Red Circles) ANSI DSL loops; (Black circles) CSA DSL loops; (Green squares) single section AWG26 cables, from $304.8~m$ ($1~kft$) to $5.5~km$ ($18~kft$) with $304.8~m$ ($1~kft$) increments.}

\subsection{The Wireless case} \label{sec.UniversalWireless}

The correlation between channel gain (path loss) and RMS-DS reported here for wireline channels has been also observed in both indoor and outdoor radio channels \cite{Deva1986, BergRupreWeck91, FeueBlacRappa1994, GhasJanaRice2004}, although a review of the literature up to a 1994 shows that this correlation has been generally ignored \cite{HashThol1994}. Interestingly, this correlation has also been found to hold in the presence of shadowing for urban macrocellular environments \cite{SousJovaDaig1994} and Greenstein et al. also conjecture that this may be true more in general thus suggesting that dispersion will tend to worsen under conditions of deep shadow \cite{GreErcYeh97}.

Notwithstanding an increasing awareness of this correlation in the wireless context, it is not uncommon to find wireless channel models that contain normalized gains so that the effects of the double hit due to simultaneous large dispersion and high attenuation are not correctly replicated. These considerations may also be extended to the MIMO context where the vast majority of the literature focuses on the multipath properties of the channel and thus carries out a normalization of the SNR out of the channel matrix. However, the correlation between path loss and multipath confirms that it may be methodologically flawed to carry out this normalization since it implies that channel multipath and available SNR are independent quantities. Recent contributions have pointed out the necessity of exercising some caution when normalizing the SNR in MIMO system performance studies \cite{SvanWall03, JensWall04, NielAnde06}.

\begin{table*}[htbp]
  \centering
  \begin{minipage}{\textwidth}
  \centering
  \caption{Summary of results presented in Sects. \ref {sec.OverviewStats} and \ref{sec.Universality}. Columns 1 and 2 report whether channel gains and RMS-DS are confirmed to be lognormally distributed by all the statistical tests listed in Sect. \ref {sec.OverviewStats}. Columns 3 and 4 report the correlation coefficients between channel gains and the RMS-DS or its logarithm.}\label{Tab.Summary}

{
\renewcommand{\arraystretch}{1.2}
\begin{tabular}{l|c|c|c|c|}
\cline{2-5}
    & $\overline{G}_{dB}$ & $\sigma_\tau$ & $\rho (\overline{G}_{dB},\sigma_\tau)$ & $\rho (\overline{G}_{dB},log(\sigma_\tau))$ \\ [0.5ex]
  \hline\hline
  Urban In-Home PLC         & Lognormal                     & Lognormal       & -0.5 & -0.6 \\
  Sub-Urban In-Home PLC     & Lognormal                     & Lognormal       & -0.4 & -0.5 \\
  Underground MV PLC        & Lognormal\footnote{~After removing outliers.}     & Lognormal$^a$     & -0.7 & -0.7 \\
  In-home coaxial           & Lognormal                     & Mixed results     & -0.4 & -0.4 \\
  In-home phone lines       & Mixed results                 & Lognormal$^a$ & -0.2 & -0.6 \\
  DSL-CSA loops             & Normal/Lognormal                     & Normal/Lognormal     & -0.95 & -0.95 \\
  DSL-ANSI loops            & Normal/Lognormal                     & Normal/Lognormal     & -0.97 & -0.95 \\ [1ex]
  \hline \hline
\end{tabular}

}
\end{minipage}
\end{table*}

\section{Implications of $\overline{G}-\sigma_\tau$ Correlation on ISI-Mitigation Techniques} \label{sec.ISImitigation}

The fact that channel RMS-DS is proportional to channel attenuation has implications on channel equalization, both in single carrier and in multi-carrier systems. Nevertheless, this aspect has been largely ignored in the literature.

Typically, channel models contain some sort of common channel gain normalization whereas dispersiveness of the channel remains unchanged. Equalizer performance is then assessed by varying noise power which \textit{artificially} varies the channel SNR. However, multipath richness and SNR do not vary independently from each other so that the choice of the right ISI-mitigation technique for a specific application may not be the correct one if the effects of normalization are not taken into account. For example, it is common practice in the literature to evaluate and compare equalizers performances on a given link for SNR values ranging several tens of dB. However, since large ISI is experienced only in highly attenuated channels (for fixed transmission power, this also implies low SNR), equalizers effectiveness on ISI-dominated channels is meaningful at low SNR values for a given link; on the other hand, the effectiveness of an equalizer on channels with low ISI is meaningful at high SNR values.

Similar considerations can be made for the case of multi-carrier systems where ISI is handled by resorting to a Guard Interval (GI). According to common design approaches, the GI of an OFDM symbol is chosen so as to ensure nearly full equalization, i.e. the GI length is set to be equal to the duration of the impulse response that includes a rather high percentage of the total impulse response energy. Such a choice for the GI ensures nearly full equalization and, thus, a very high Signal-to-Interference power Ratio (\textit{SIR}) - where interference is here defined as the sum of the contributions of ISI and Inter-Carrier Interference (ICI)\footnote{$^)$~In OFDM systems with insufficient GIs, ICI power is the same as ISI power if the channel duration extending beyond the GI is small compared to the OFDM symbol duration $NT_S$ \cite{NguyKuch2002}. Thus, in the following, we will generically refer to ``interference'' as the sum of the ISI and ICI components.}. Longer GIs for the same OFDM symbol duration imply lower transmission efficiency and this can be compensated by increasing the number of employed sub-carriers (for the same bandwidth). Thus, full equalization in OFDM always implies either a decrease of the data rate due to lower transmission efficiency or an increase in circuit complexity due to the increase of the number of carriers.

If the GI is not long enough, then the received signal is affected by both ISI and ICI. Since data symbols in two adjacent OFDM symbols are statistically independent, the total (ISI plus ICI) interference power is the sum of the ICI and ISI powers. Therefore, the total received power $P_R(k)$ on the $k$-th sub-channel at the output of the DFT can be written as ($M$ is the number of sub-carriers, $\nu$ the number of samples in the CP):
\begin{eqnarray} \label{eq.PowerComponents}
  P_R(k) &=& \frac{M}{M+\nu} P_T |H_k|^2 + N_0 \\
         &=& \frac{M}{M+\nu} P_T \left( P_U(k) + P_I(k) \right) + N_0
\end{eqnarray}
where $P_U(k)$ and $P_I(k)$ denote the power contributions (on the $k$-th subcarrier) to the useful signal and to the interference, respectively. Recalling that both ICI and ISI powers are independent of the sub-carrier index and functions of $M$ and $\nu$ \cite{NguyKuch2002}, we can write the following Signal to Noise plus Interference Ratio (SNIR):
\begin{eqnarray} \label{eq:SNRandSIR}
  \textit{SNIR}(M,\nu) &=& \frac{\frac{M}{M+\nu} P_T P_U(M,\nu)} {\frac{M}{M+\nu} P_T P_I(M,\nu) + N_0} \\
                       &=& \frac{\frac{M}{M+\nu} P_T P_U(M,\nu)} {\frac{M}{M+\nu} P_T (\overline{G}-P_U(M,\nu)) + N_0}
\end{eqnarray}
where we have dropped the dependency of $k$ and explicitly shown the dependency on $M$ and $\nu$.

The expression of the achievable bit rate $C_{MC}(M,\nu)$ for the case of a partially equalized multi-carrier system is a logarithmic function of \eqref{eq:SNRandSIR} and, thus, will also depend on $M$ and $\nu$. As the CP decreases, transmission efficiency $M/(M+\nu)$ will increase while $P_U(M,\nu)$ will decrease and $P_I(M,\nu)$ will increase as more ISI and ICI will be present. As the CP increases, transmission efficiency will decrease while $P_U(M,\nu)$ will grow towards the channel gain $\overline{G}$ and $P_I(M,\nu)$ will decrease to zero. Clearly, for a channel with a given gain, RMS-DS and $N_0$, there exists an optimal $(M,\nu)$ pair that maximizes the achievable data rate. When $N_0$ is high, more channel distortion (shorter CP, larger $P_I(M,\nu)$) can be allowed as long as $M/(M+\nu)P_T P_I(M,\nu) \ll N_0$ as this would increase achievable data rate due to the increase of the transmission efficiency $M/(M+\nu)$; on the other hand, if $N_0$ is small, little channel distortion (longer CP, smaller $P_I(M,\nu)$) can be allowed since a large $M/(M+\nu)P_T P_I(M,\nu)$ would cause a drop in the achievable SNIR.

The above considerations confirm that the achievable data rate depends on SNIR through $\overline{G}$ and $\nu$, and $\nu$ is typically chosen on the basis of $\sigma_\tau$ that is correlated to $\overline{G}$. Thus, the capability of a channel model to replicate this correlation is important for the correct assessment of the performance of a multi-carrier scheme.

\begin{remark}
The negative correlation between RMS-DS and channel gains suggest that full equalization in multi-carrier systems is indeed a sub-optimal choice when the channel is noise limited as pointed out in \cite{Link:Galli09keynoteISPLC}. The immediate advantage of pursuing partial equalization is that transmission efficiency increases without the need to increase the number of carriers which raises complexity - see also Sect. IV.B in \cite{Galli2009lognormal}. This conclusion is in line with \cite{SeoWilGel97} where the Authors noted that sometimes there is an advantage in letting some taps ``escape'' the GI, especially when the probability of having a long channel response is small. The fact that the probability that a channel has a large RMS-DS is directly proportional to the probability that a channel has a large attenuation becomes very important in noise limited channels like the PLC one. In fact, the data rate increase gained by shortening the GI is ``protected'' by the fact that noise is generally larger than ISI/ICI and that \textit{SNR} and \textit{SIR} decrease simultaneously when links characterized by both high attenuation \textit{and} large RMS-DS are encountered. This result has also been recently confirmed on the basis of simulations results by Tonello et al. \cite{ToneDaleLamp2010CPadaption}.
\end{remark}

\section{The Proposed Channel Modeling Approach}\label{sec.GeneralModel}

The proposed statistical channel modeling approach starts with defining the desired impulse response $h^\star(t)$ as:
\begin{equation}\label{eq.GeneralChannelModel}
    h^\star(t) = \sum_{k=0}^{L-1} h^\star[k] \delta(t-k \tau^\star),
\end{equation}
where taps are equi-spaced and $\tau^\star$ is not necessarily equal to the sampling time $T_S$. The reason for using a delay different from $T_S$ is to allow for imposing a specific RMS-DS on the generated impulse response.

The impulse response $h^\star(t)$ is the physical baseband channel response. In practice, most modems do not operate over baseband but in passband. For example, broadband communication over PLs typically occurs in the frequencies above 2 MHz, communications over PH occurs above voice band, and communications over CX is often at RF. The equivalent impulse channel response $h_{eq}^\star(t)$ seen at the receiver is given by the convolution of the physical channel response with the transmit pulse shaper and the receive anti-aliasing filter. Assuming that the cascade of transmit and receive filters has a Nyquist characteristic $p(t)$, we can write the equivalent impulse response $h_{eq}^\star(t) = h^\star(t) \star p(t)$ directly in the discrete time as follows ($k=0, 1,\ldots, L-1$):
\begin{equation} \label{eq.EquivImpulseresponseDT}
  h_{eq}^\star[k] = \int_{\infty}^{\infty} h^\star(\tau) p(kT_S-\tau) d\tau
\end{equation}
where $p(t)$ is the low pass equivalent kernel of the passband filter which can be easily obtained by taking the Inverse Fourier Transform of the corresponding frequency response down-shifted by the carrier frequency.

The above expression holds for LTI channels, but not for LTV channels as the LTV physical channel and the LTI receive filter kernels cannot be swapped. For the LTV channel case, we then have:
\begin{equation}\label{eq.tv-rel11}
h_{eq}^\star[k,l]=\int_{-\infty}^{\infty} \int_{-\infty}^{\infty} h^\star(kT_s-\xi,\tau) p\prime(lT_s-\tau-\xi) p\prime\prime(\xi) d\tau d\xi,
\end{equation}
where $p\prime(t)$ is the transmit pulse shaping filter, $p\prime\prime(t)$ is the receive filter, $p(t)=p\prime(t) \star p\prime\prime(t)$ and $h^\star(t,\tau)$ is the response of the LTV channel.

Channel time-variance is of practical interest here since the PLC channel has been found to be a Linear and Periodically Time Varying (LPTV) channel \cite{CanDieCor2002, Cav2004, CanCorDie2006, SunScaGal2008}. An LPTV channel with period $f_0=1/T_0$ is such that $h(t+kT_0,\xi)=h(t,\xi), \forall k\in {\mathbb Z}$. Although the modeling approach proposed here is limited to the generation of LTI channel realizations, it can still be used to generate LPTV channel realizations since LTI models are indeed sufficient to generate LPTV realizations using, for example, the LPTV Zadeh decomposition \cite{Zadeh1950}. More in detail, Zadeh introduced the following expansion valid for LPTV systems:
\begin{equation}\label{eq.ctmodel-chanFS}
h(t,\tau)=\sum_{m=-\infty}^{+\infty} h_m(\tau)e^{j \frac{2\pi m}{ T_0}t}
\end{equation}
where $h_m(\tau)$ are the so-called $harmonic$ impulse responses:
\begin{equation}\label{eq.ctmodel-chanFS2}
h_{m}(\tau)=\frac{1}{T_0}\int_{0}^{T_0} h(t,\tau)e^{-j2\pi m f_0 t} dt
\end{equation}
In Zadeh's expansion, an LPTV channel is equivalently represented as a bank of LTI channels whose outputs are modulated by Fourier harmonics with frequencies that are integer multiples of the fundamental frequency $1/T_0$. Estimators for $h_{m}(\tau)$ can be found in \cite{CanCorDie2006}, \cite{GallScagXXlptv}.

We can now give the procedure for generating LTI channel realization according to the proposed statistical channel modeling approach:

\begin{enumerate}

  \item Impose the Power-Delay Profile (PDP) shape by choosing $L$ and setting channel tap amplitudes $h^\star[k],~(k=0,1,\ldots,L-1)$ according to a specific criterion, e.g. at random, exponentially, equi-powered, etc. Channel taps can be positive, negative, real or complex.

  \item Extract $\overline{A}_{dB}^\star=-\overline{G}_{dB}^\star$ from a lognormal distribution appropriate for the scenario being simulated, e.g. using the values in Tables \ref{tab.PLCstats}-\ref{tab.ALLstats}.

  \item Compute the RMS-DS value $\sigma_\tau^\star$ corresponding to the extracted value of $\overline{A}_{dB}^\star$ as follows:
  \begin{enumerate}
    \item If the RMS-DS kurtosis of the considered scenario is comparable to or lower than $3$ (the kurtosis of a Gaussian distribution) then use the regression equation in \eqref{eq.PLCrobregr} or \eqref{eq.PLCrobregrLOG}.
    \item If the RMS-DS kurtosis of the considered scenario is larger than $3$, then extract a random variable from the conditional distribution - e.g. see the one introduced in Sect. \ref{sec.RMSDSdef} for the IH sub-urban case.
  \end{enumerate}

  \item Normalize $h^\star(t)$ so that $10log(\sum_k |h^\star[k]|^2) = \overline{G}_{dB}^\star$.

  \item Set $\tau^\star$ in \eqref{eq.GeneralChannelModel} so that $h^\star(t)$ has an RMS-DS equal to the desired $\sigma_\tau^\star$ set in Step 2 by doing as follows:

  \begin{enumerate}
      \item Compute the RMS-DS $\sigma_{T_S}$ of $h^\star(t)$ in \eqref{eq.GeneralChannelModel} when posing $\tau^\star=T_S$;

      \item Set $\tau^\star$ in the model of \eqref{eq.GeneralChannelModel} as follows:

\begin{equation}
          \tau^\star=\frac{\sigma_\tau^\star}{\sigma_{T_S}}T_S
\end{equation}

 \end{enumerate}

 \end{enumerate}

Step 2 exploits the fact that all investigated wireline channels exhibited lognormal channel gains. Although in Sect. \ref{sec.Universality} we reported that in some cases there was an uncertainty in the determination of the distribution of RMS-DS, this uncertainty can be by-passed by avoiding to extract randomly the target RMS-DS $\sigma_\tau^\star$ and by setting it using the robust regression line as directed in Step 3a).

In Step 3 above, we have taken into account that the regression equation \eqref{eq.PLCrobregr} may not be accurate in certain scenarios. The choice between Steps 3.a) and 3.b) should be made on a case-by-case basis. For example, for strongly leptokurtic distributions as the US sub-urban one, the regression line is not accurate at low channel gains (see Figure \ref{fig.PLCScatter}); in this case, it would be better to draw $\sigma_\tau$ from a lognormal distribution with parameters as in Table \ref{tab.ALLstats} when the value of the channel gain $\overline{G}_{dB}$ is lower than a certain thershold. As the scatter plot in Figure \ref{fig.PLCScatter} confirms, the regression equation in \eqref{eq.PLCrobregr} is accurate enough for computing the RMS-DS corresponding to a given channel gain for the US urban case.

Since the correlation between channel gains and RMS-DS is a property shared by several wireline channels, the channel modeling approach presented here can be used in a wide variety of cases. Its capability of replicating this physical property allows for a more realistic assessment of equalizers' effectiveness or multi-carrier's design, something not always ensured by other standardized channels models which contain some sort of SNR normalization that allows channels with the same normalized gain to exhibit wide variations of RMS-DS.

In Figure \ref{fig.ExampleRandom}, we show two realizations of IH PL channels obtained using the above procedure. These two realizations have been obtained using Guassian distributed channel taps $h^\star[k]$ and similar results can be obtained using other distributions for the tap amplitude, e.g. uniform, lognormal, etc.

\inserttwofigsV{8.8cm}{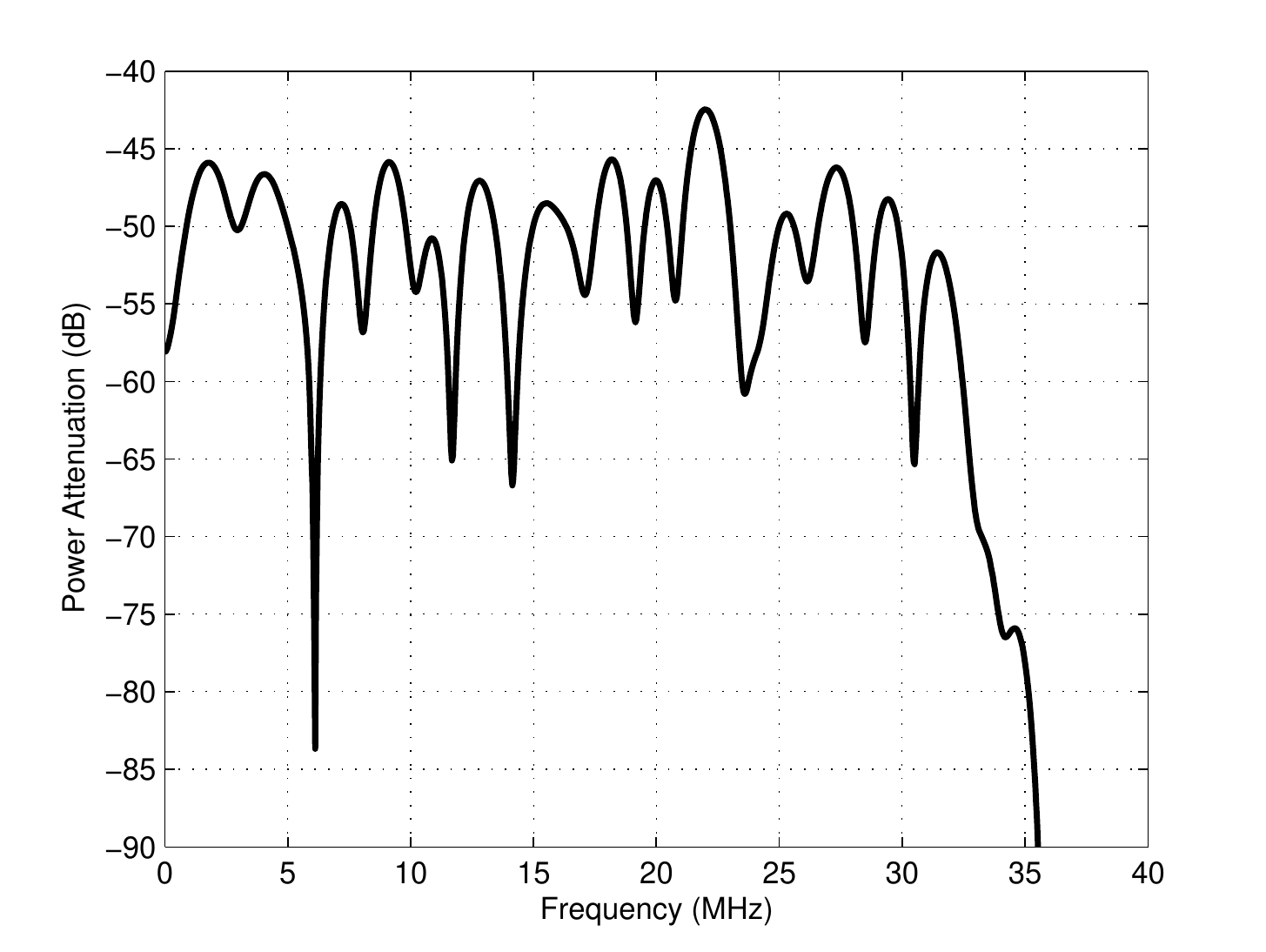}{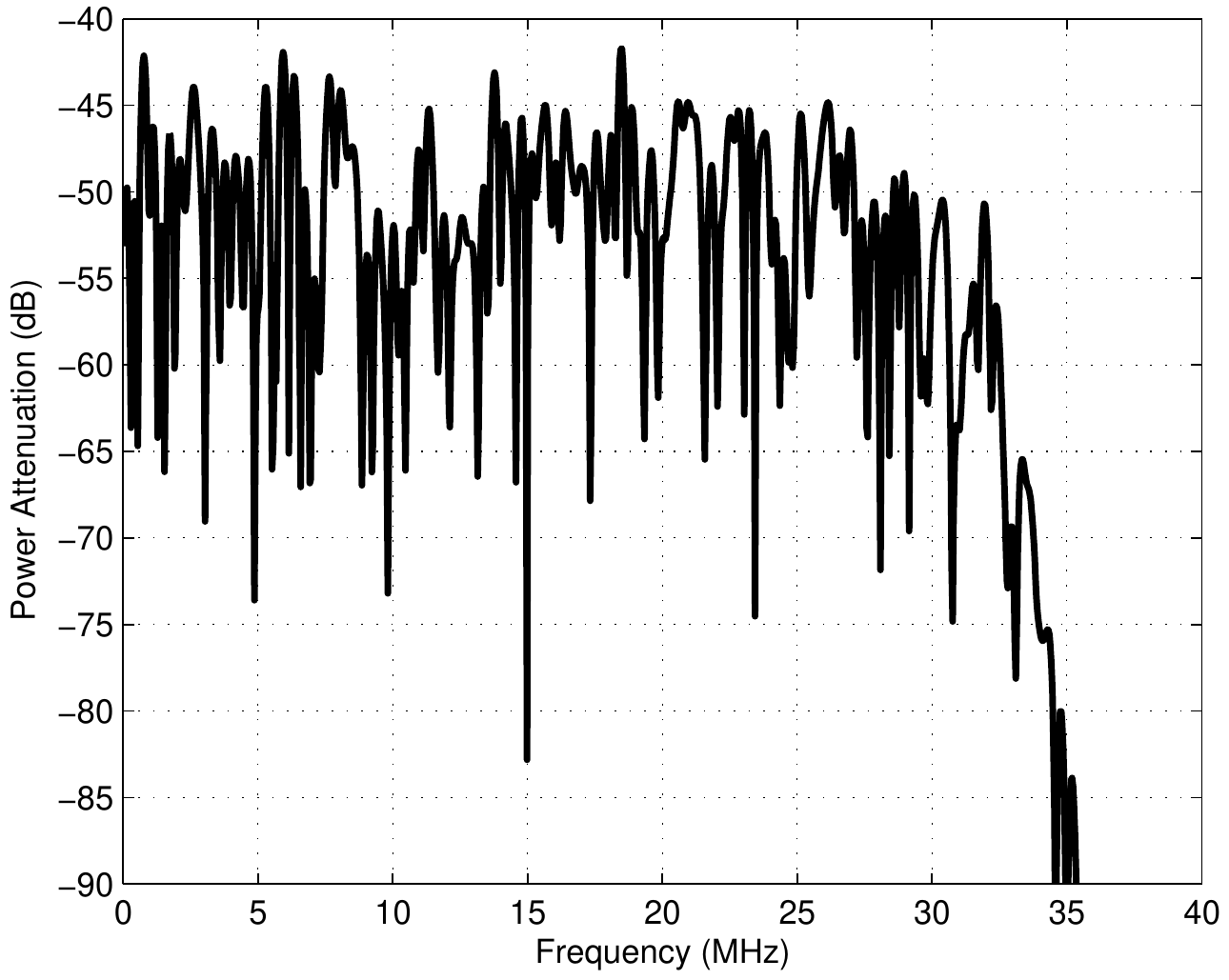}{\label{fig.ExampleRandom}}
{Examples of channel realizations in the [0-30] MHz band obtained using the proposed channel model for the case of randomly generated Gausssian distributed taps and after raised cosine filtering with roll off equal to 0.2: (a) $\overline{G}_{dB}^\star=-50~dB$, $\sigma_\tau^\star=0.2~\mu s$, and $L=50$ taps. (b) $\overline{G}_{dB}^\star=-50~dB$, $\sigma_\tau^\star=1.3~\mu s$, and $L=1000$ taps.}

\subsection{A Simple Two-Tap Channel Model} \label{sec.SimplifiedModel}
The modeling approach proposed here allows us to obtain channel realistic channel realizations. It is certainly appealing to define a very simple statistical channel model as a particular case of the general procedure given earlier. This can be done choosing a simple model with two equi-amplitude taps whose amplitude and differential delay yield to channel realizations where gains and RMS-DS are correlated random variables. Channel models based on a two-path model are not new as they have been used in wireless TDMA standards (IS-54, IS-136, GSM) or for the High Frequency (HF) radio channel. These models specify two equal-power independently faded rays spaced by a fixed delay where delay values are tabulated values that mimic specific link conditions. This simple two-tap wireless modeling allowed easy comparison between competing schemes. The novelty of the modeling approach proposed here is that tap amplitude and differential delay are not fixed, tabulated, and independent values but they are \textit{correlated random variables}.

An interesting question that naturally arises is the following: how important the specific PDP shape of the channel is? An important result by Glance and Greenstein \cite{GlaGre83} states that, if the RMS-DS of a channel is less than 20\% of the symbol duration time, the dispersion effects of the channel (in terms of BER vs SNR) are fully characterized by the RMS-DS alone and are independent of the specific PDP shape. On the other hand, if the RMS-DS is larger than 20\% of the symbol period, then the effects of dispersion are dependent on the specific PDP shape. This is relevant to the case addressed here since the RMS-DS of wireline channels is always less than 20\% of the symbol duration time\footnote{$^)$~In the case of DSL, this is true only when channel shortening is performed at the receiver and this is usually the norm.}.

The underlying assumptions made in \cite{GlaGre83} are: single carrier narrowband links, low order modulations (BPSK, QPSK), ISI present at the receiver (partial or no equalization performed). Under these assumptions, the actual shape of the PDP is then not critical for the performance evaluation of some single carrier wireline standards, e.g ISO/IEC 14908 (Loneworks), ISO/IEC 14543 (KNX), CEA-600.31 (CEBus) and ITU-T G.9954. However, in the past several years, industry has converged towards adopting multi-carrier schemes for wireline channels, e.g. TIA-1113 (HomePlug 1.0), IEEE 1901 Broadband over Power Lines\footnote{$^)$~The IEEE 1901 Draft Standard specifies two PHY/MACs (one based on HomePlug's FFT-OFDM \cite{AfkhKataYong2005homeplugAV} and one based on Panasonic's Wavelet-OFDM \cite{GalKogKod2008}) and a coexistence scheme called Inter-system Protocol (ISP) \cite{GalKurOhu2009}.} \cite{GalLog2008}, and ITU-T Recommendation G.9960/G.9961 (G.hn) \cite{OksGal2009} - for an updated overview of all PLC standards see \cite{Book:PLC-NEW-2010Chap7}. In the multi-carrier case, the actual PDP shape may be sometimes relevant to performance evaluation and this depends on the type of bit loading adopted, CP length, and so on. In those cases where the PDP shape is indeed relevant, then the proposed approach is still suitable for a comparative analysis of communication schemes and any generalization of results would of course require testing against multiple PDP families - which can still be done using the five-step approach outlined earlier.

Let us now consider the simplest possible ISI-channel model, i.e. a two-tap channel. In this case, gain and RMS-DS are given by the following expressions:
\begin{eqnarray}\label{eq.2TAPgain}
  \overline{G}_{dB} &=& 10\log(\overline{G}) = 10\log (\lvert h_1 \rvert^2 + \lvert h_2 \rvert^2) \\ \label{eq.2TAPrmsds}
  \sigma_\tau       &=& \frac{|h_1 \cdot h_2|}{\lvert h_1 \rvert^2 + \lvert h_2 \rvert^2} \tau
\end{eqnarray}
Thus, in the simplest case of a two-tap equi-powered channel, the last two steps 4) and 5) of the general channel generation procedure given earlier can be replaced with the following simplified steps:
\begin{description}
  \item[4$^\prime)$] Set the energy of the two channel taps as $|h^\star|^2=|h_1|^2=|h_2|^2=0.5\cdot10^{\overline{G}_{dB}^\star/10}$.

  \item[5$^\prime)$] Set the differential delay in the two-tap model as $\tau^\star = 2 \sigma_\tau^\star$.
\end{description}
The resulting two-tap channel model would then be:
\begin{equation}\label{eq.FinalChannelModel}
    h^\star(t) = h^\star [ \delta(t) + \delta(t-\tau^\star)]~\Leftrightarrow~H(f) = h^\star [1 + e^{-j 2 \pi f \tau^\star}]
\end{equation}
Although simple, a two-tap channel generated using the proposed approach is able to generate channels with strong frequency selectivity such that sub-carriers experience a wide range of channel attenuation values. On the other hand, using only two channel taps introduces periodic features in the transfer function for values of RMS-DS above a certain threshold.

\insertonefig{8.8cm}{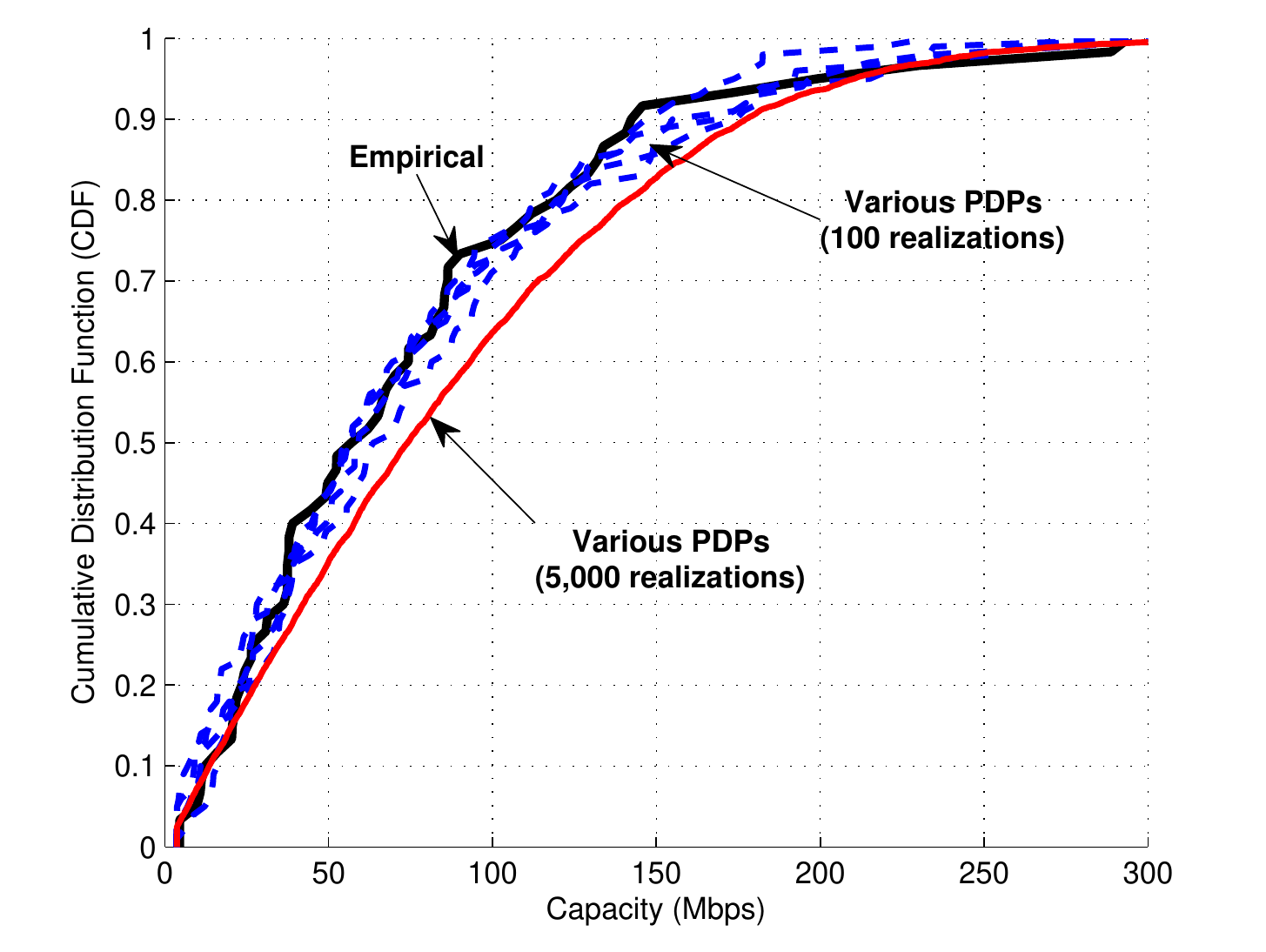}{\label{fig.HPAcap}}
{Capacity CDF curves calculated for the US sub-urban IH PLC case. Solid bold black: empirical data. Solid bold red: 5,000 channel realizations with a two-tap equi-amplitude PDP or a randomly generated PDP with $L>4$ (curves overlap). Dashed blue: 100 channel realizations with various PDPs and $L \geq 2$ (random, two-tap, etc.).}

\insertonefig{8.8cm}{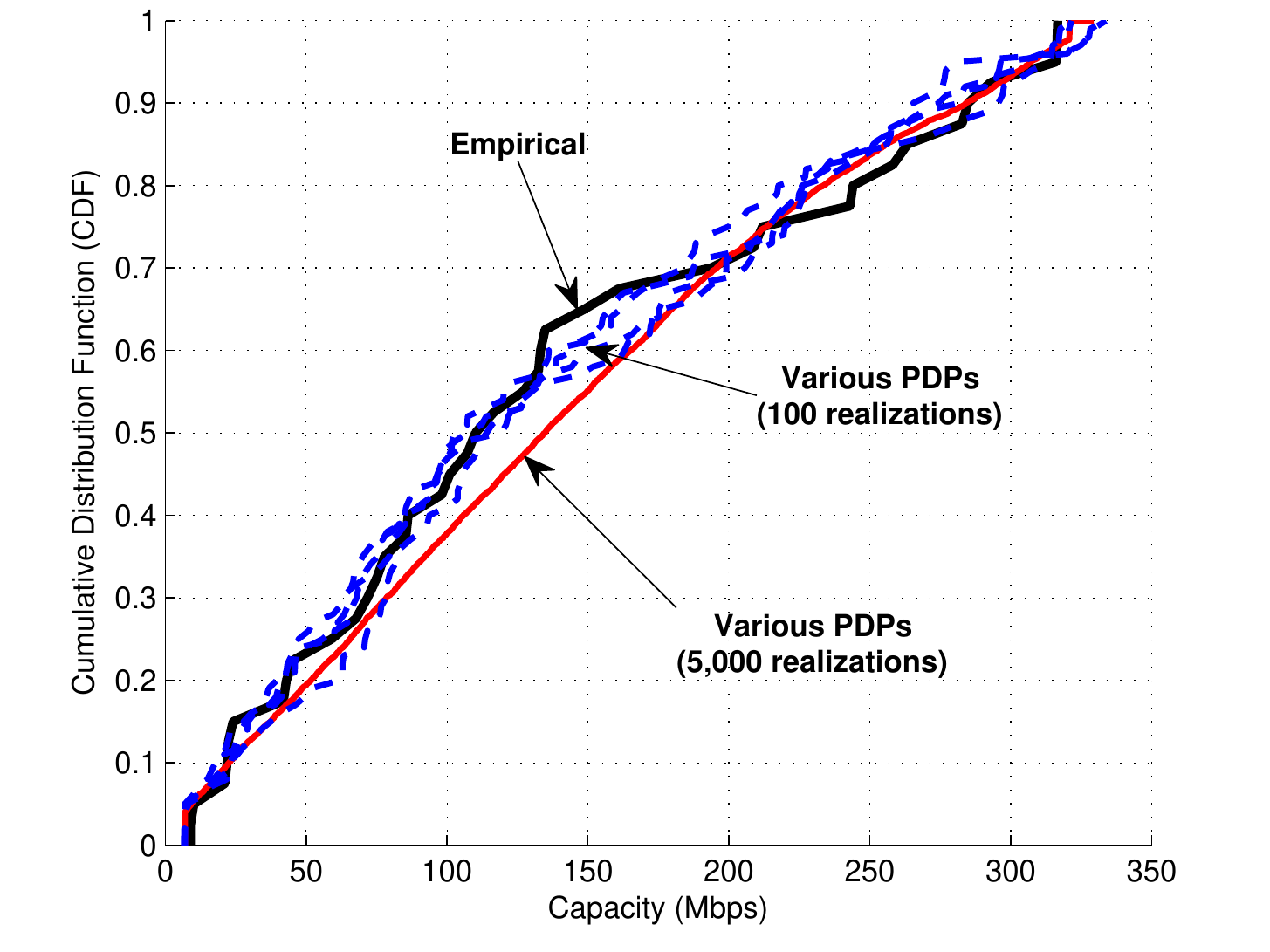}{\label{fig.PanaUScap}}
{Same as in Figure \ref{fig.HPAcap}, but for the case of US urban homes.}

\section{Capacity and Coverage Curves} \label{sec.Simulations}
In this Section we will show how the proposed statistical channel model yields results that are close to the ones obtained through measurements. We consider here the PLC case because of the availability of empirical data.

In a first set of simulations, we have compared the empirical Cumulative Distribution Function (CDF) of channel capacity with the CDF obtained when channel realizations are generated using the proposed statistical channel model. The CDF plots of capacity are useful to calculate the coverage that can be achieved in a house, i.e. the percentage of outlet pairs that are able to support a certain data rate. Achievable capacity can be expressed as:
\begin{equation}\label{eq:capacity}
    C = W \sum_{k \in \cal K} log_2 \left( 1+\frac{\textit{SNR}_k}{\Gamma} \right)~~~~~bits/sec
\end{equation}
where $k$ is the sub-channel index, $\cal K$ is set of sub-channels carrying information, $W$ is the utilized frequency band, and $\Gamma$ represents an SNR gap factor from Shannon capacity that accounts for the deployment of practical modulation and coding schemes \cite{Book:StaCioSil1999}. This gap measures the efficiency of the transmission scheme with respect to the best possible performance in Additive White Gaussian Noise (AWGN) and can be approximated in dB as follows: $\Gamma_{dB} \approx 9.8 + \gamma_m - \gamma_c$, where $\gamma_c$ is the the coding gain in dB ensured by the employed line code and $\gamma_m$ is the desired system margin in dB.

For the simulations, we have considered: a bandwidth $W=28 MHz$ as today's broadband PLC devices operate in $[2-30]$ MHz; a transmit power density of $P_T=-55$ dBm/Hz; a white noise power density of $N_0=-120$ dBm/Hz\footnote{$^)$~Noise in PLC is not AWGN, and is often impulsive, colored and non-Gaussian \cite{GotRapDos2004}. To account in part for those components, we have chosen a higher than usual noise power density $N_0$.}; a gap of $\Gamma_{dB}=7$ dB. In calculating the gap, we have considered a coding gain of $8.8$ dB \cite{Galli2010fec} and a classical $6$ dB margin. A practical spectral efficiency cap of $12$ bits/sec/Hz was also imposed.

CDF plots are shown in Figures \ref{fig.HPAcap} and \ref{fig.PanaUScap} for the two cases of US urban and sub-urban IH PLC, respectively. In the sub-urban case, there is a significant discrepancy between the empirical CDF and the CDF obtained using 5,000 channel realizations. However, when a smaller sample of channel realizations is used, the statistical model is able to generate a family of CDF curves that include CDFs very close to the empirically observed one. Thus the discrepancy observable in Figure \ref{fig.HPAcap} should not be attributed to an incapability of the proposed channel model to replicate empirical results but to the fact that most likely the collected measurements represent a pessimistic data set. On the other hand, the urban measured data set is very close to the one obtained via simulation. Interestingly, the CDF obtained using 5,000 channel realizations with $L>4$ randomly generated taps coincides with the CDF obtained using 5,000 two-tap equi-powered channel realizations.

As is well known, capacity strongly depends on the SNR available at the receiver. We have verified that capacity varies linearly with channel gains (path loss). Figure \ref{fig.CAPvs}.(a) clearly confirms the linear trend for the case of the IH PLC channel with measured correlations of 0.96 and 0.98 for the urban and sub-urban cases, respectively. Since there is a correlation between channel gains and RMS-DS, there is also a correlation between SNR and RMS-DS when fixed transmit power is considered. Thus, it should not be surprising that capacity should also depend (indirectly) on channel RMS-DS. This is confirmed by the plot in Figure \ref{fig.CAPvs}.(b) where a marked trend of inverse proportionality between capacity and RMS-DS can be observed. These results suggest that channels characterized by multipath richness (large RMS-DS) are also characterized by low channel capacity. This dependency of channel capacity on RMS-DS is reported here for the first time.

\inserttwofigsV{8.8cm}{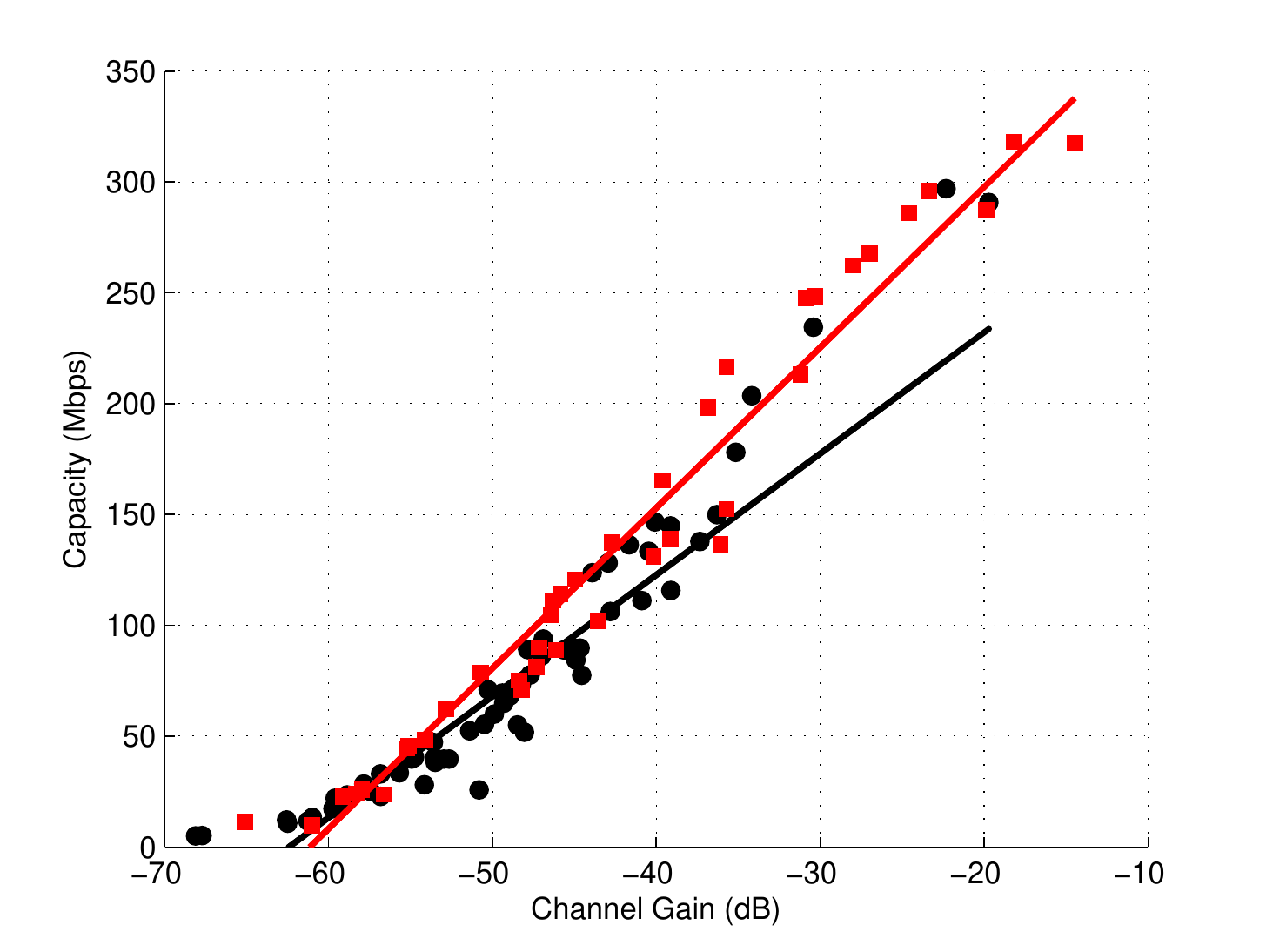}{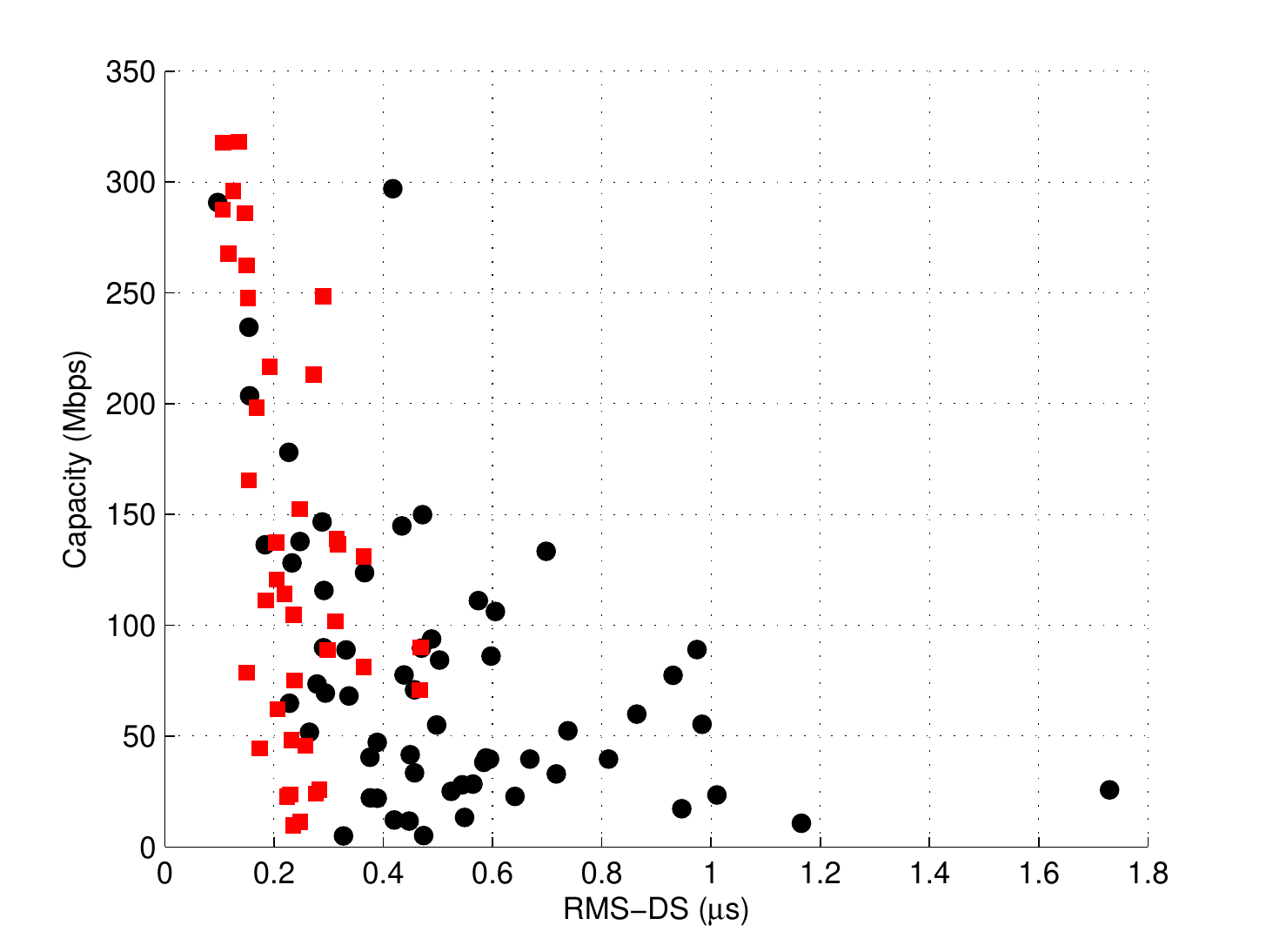}{\label{fig.CAPvs}}
{Capacity versus channel gain $\overline{G}_{dB}$ (a) and versus RMS-DS (b) for two cases of measured IH PLC channels: US sub-urban (black circle) and US urban (red square).}

\section{Conclusions}\label{sec.Conclusions}

The transfer function of a TL-based channels can be accurately and deterministically calculated once the link topology is known in a way that is similar to (albeit much easier than) the wireless case (ray tracing). However, the variability of link topologies and wiring practices gives rise to a stochastic aspect of TL-based channels that has been seldom accounted for in the literature.

We presented a novel and yet simple approach to the modeling of TL-based channels that can be used for the comparative analysis of modulation and coding schemes. The fundamental feature of the proposed statistical model is that it mimics a fundamental property of several wireline channels: the negative correlation between channel gain and RMS-DS. The proposed statistical modeling approach can be employed to simulate a wide variety of wireline channels and allows generating any channel power-delay profile including a convenient and simple two-tap one. Furthermore, the proposed model allows accurate capacity and coverage analysis, multi-carrier parameter optimization, appropriate interpretation of equalization performances, and also lends itself to an extension that allows modeling LPTV channels such as the PLC one.

Specific scenarios can be simulated by setting the parameters of the model on the basis of empirical data obtained by grouping measurements according to a specific criterion, e.g., on the basis of country, on the basis of house type, etc. Defining specific scenarios on the basis of statistical information would allow easy replication of results and comparative analysis of communications techniques - something that today is still difficult in the PLC context because of the lack of a commonly agreed upon statistical channel model.

Since sharing of statistical averages is often much easier than exchanging actual impulse response measurements, we believe that the proposed model can become a reference channel model for evaluating communications schemes over a wide variety of wireline physical channels.

\section*{Acknowledgment}
The Author wishes to express his gratitude to Nobuhiko Noma and Hisao Koga of Panasonic for taking the US urban power line channel measurements and to Brent Zitting of IBEC for generously sharing the underground MV power line data. Special thanks also to Kate Wilson and Larry Greenstein for their very valuable feedback that helped improving the paper.

\bibliographystyle{IEEEtran}
\bibliography{IEEEabrv,All_Papers,PLC_Papers}

\begin{IEEEbiography}[\addphoto{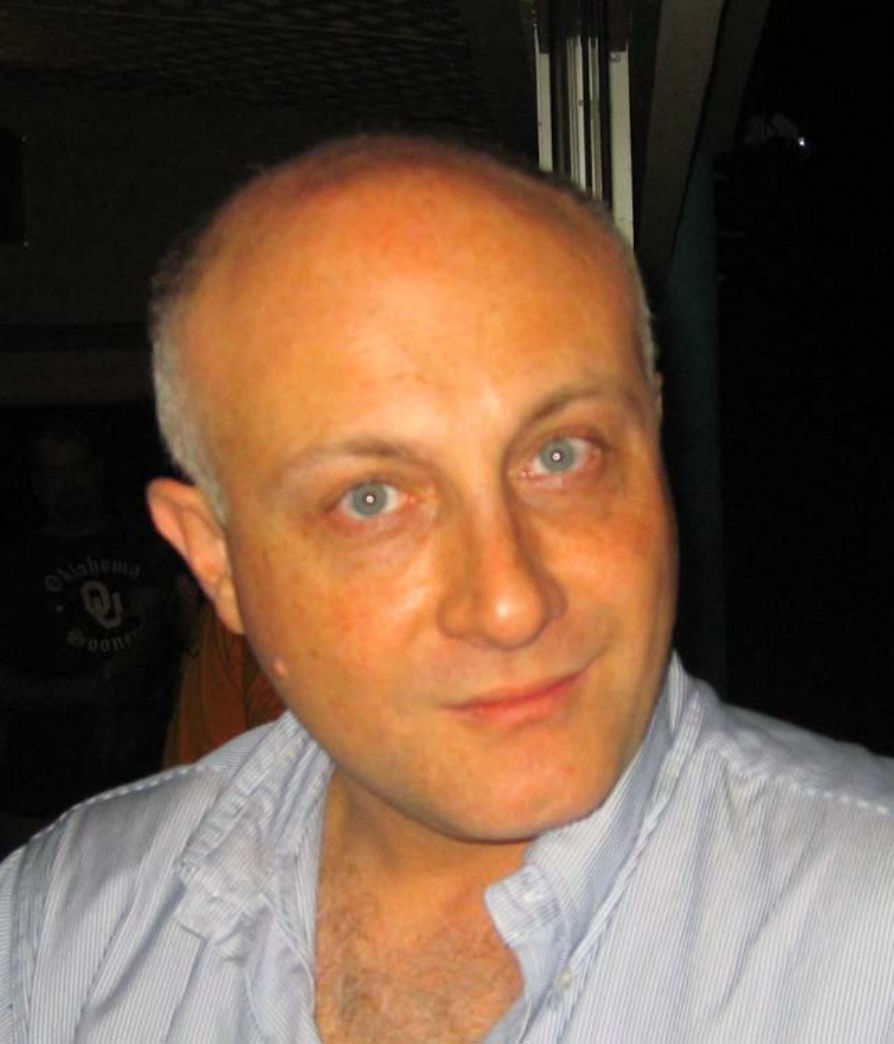}]
{Stefano Galli} (S'95, M'98, SM'05) received his M.Sc. and Ph.D. degrees in Electrical Engineering from the University of Rome ``La Sapienza'' (Italy) in 1994 and 1998, respectively. Currently, he is the Director of Technology Strategy of ASSIA where he leads the overall standardization strategy of the company and contributes to the company's efforts in the areas of wired/wireless access and home area networking. Prior to this position, he was in Panasonic Corporation from 2006 to 2010 as Lead Scientist in the Strategic R\&D Planning Office and then as the Director of Energy Solutions R\&D. From 1998 to 2006, he was a Senior Scientist in Bellcore (now Telcordia Technologies).\\
Dr. Galli is involved in a variety of capacities in Power Line Communications and Smart Grid activities. He currently serves as Chair of the PAP-15 Coexistence subgroup instituted by the US National Institute of Standards and Technology (NIST), Chair of the IEEE Communications Society (ComSoc) Smart Grid Communications Committee, Member-at-Large of the IEEE ComSoc Board of Governors, Member of the Energy and Policy Committee of IEEE-USA, and Editor for the IEEE Transactions on Smart Grid and the IEEE Transactions on Communications (Wireline Systems and Smart Grid Communications). He is also the founder and first Chair of the IEEE ComSoc Technical Committee on Power Line Communications (2004-2010), and the past Co-Chair of the ``Communications Technology'' Task Force of the IEEE P2030 Smart Grid Interoperability Standard (2009-2010). \\
Dr. Galli has worked on numerous wireless/wired communications technologies and is now focusing on Power Line Communications and Smart Grid. He is an IEEE Senior Member, holds fourteen issued and pending patents, has published over 90 peer-reviewed papers, has submitted tens of standards contributions, and has received the 2010 IEEE ISPLC Best Paper Award.
\end{IEEEbiography}

\balance

\end{document}